\documentclass[11pt,a4paper]{article} 
\pdfoutput=1
\usepackage{jheppub}

\usepackage{amsmath,amssymb,epsfig,bm,amsfonts,mathrsfs,yfonts}
\allowdisplaybreaks

\begin{document}

\title{Statistics of initial density perturbations in heavy ion collisions and their fluid dynamic response}

\author{Stefan Floerchinger and Urs Achim Wiedemann}
\affiliation{Physics Department, Theory Unit, CERN,\\
CH-1211 Gen\`eve 23, Switzerland}

\emailAdd{Stefan.Floerchinger@cern.ch}
\emailAdd{Urs.Wiedemann@cern.ch}

\abstract{An interesting opportunity to determine thermodynamic and transport properties in more detail
is to identify generic statistical properties of initial density perturbations. 
 Here we study event-by-event fluctuations in terms of correlation functions for two models that can be solved analytically. The first assumes Gaussian fluctuations around a distribution that is fixed by the collision geometry but leads to non-Gaussian features after averaging over the reaction plane orientation at non-zero impact parameter. In this context, we derive a three-parameter extension of the commonly used Bessel-Gaussian
event-by-event distribution of harmonic flow coefficients. Secondly, we study a model of $N$ independent point
sources for which connected $n$-point correlation functions of initial perturbations scale like
$1/N^{n-1}$. This scaling is violated for non-central collisions in a way that can be characterized by its
impact parameter dependence. We discuss to what extent these are generic properties that can be 
expected to hold for any model of initial conditions, and how this can improve the fluid dynamical 
analysis of heavy ion collisions. 
}
% \preprint{CERN-PH-TH-2014-092}

\maketitle

\section{Introduction}
In recent years, data from the LHC~\cite{ALICE:2011ab,Chatrchyan:2012ta,ATLAS:2012at} and 
RHIC~\cite{Adare:2011tg,Adamczyk:2013gw} have given strong support to the paradigm that
the QCD matter produced in ultra-relativistic heavy-ion collisions evolves like an almost perfect fluid
(for reviews, see refs.~\cite{Heinz:2013th,Gale:2013da,Hippolyte:2012yu,Teaney:2009qa}).
Hadronic spectra and particle correlations at low transverse momentum can be understood 
as the fluid dynamic response to fluctuating initial conditions~\cite{Alver:2008zza,Alver:2010gr}
(see also refs.~\cite{Mishra:2007tw,Broniowski:2007ft,Sorensen:2008zk,Takahashi:2009na}). 
This is by now supported by a large number of detailed studies~\cite{Qiu:2011iv,Schenke:2011bn,Bhalerao:2011yg,Schenke:2012wb,Gale:2012rq,Shen:2012vn,Holopainen:2010gz,Teaney:2010vd,Teaney:2012ke,Gardim:2011xv,Petersen:2012qc,Qian:2013nba,Niemi:2012aj,Deng:2011at,Floerchinger:2013rya,DelZanna:2013eua}. Since dissipative QCD hydrodynamics 
can be formulated entirely in terms of quantities that are calculable from first principles in finite temperature 
QCD, the observed fluid dynamic behaviour is at the basis of connecting measurements in these strongly
evolving mesoscopic systems to properties of QCD thermodynamics. 

In practice, testing QCD thermodynamics experimentally is complicated by the fact that data result 
from a convoluted time history that depends not only on hydrodynamic evolution but also on initial
conditions and hadronization. In particular, essentially all flow measurements are correlation measurements
and correlations can be present already in the initial conditions, or they can arise (or be attenuated) 
dynamically during the hydrodynamic evolution or during hadronization, respectively. 
The theory framework for determining the dynamical evolution is clear: once the thermodynamic
information and transport properties entering the equations of dissipative fluid dynamics are specified, the propagation of
a known initial condition can be controlled. A significant number of tools and techniques has been developed to this end~\cite{Romatschke:2009kr,Schaefer:2014xma}.  However, the initial conditions are arguably less controlled so far.  Understanding
their dynamical origin and their statistical properties is now becoming a major focus of current research. 

Earlier efforts in this direction have focussed mainly on exploring the density distributions generated
in Monte Carlo models that implement variants of the optical Glauber model~\cite{Miller:2007ri,Alver:2008aq,Broniowski:2007ft,Broniowski:2007nz,Blaizot:2014wba} or supplement these
with effects from parton saturation physics~\cite{Schenke:2012wb,Dumitru:2012yr,Rybczynski:2013yba,Bzdak:2013zma}, or in dynamically more complete code-based formulations~\cite{Avsar:2010rf,Flensburg:2011wx}. In a different direction, the simplifying assumption that initial density perturbations
follow a Gaussian distribution has served since long as a baseline for characterizing initial 
conditions~\cite{Voloshin:2007pc,Floerchinger:2013vua,Blaizot:2014nia}. The question arises
to what extent such studies explore only a possibly limited range of the total parameter space of conceivable initial conditions, or whether it is possible to identify universal features that any phenomenologically relevant model of initial conditions is expected to satisfy on general grounds.
One such general consideration that applies to heavy ion collisions is that particle production arises
from a large number of essentially independent sources with identical statistical properties.  
It is well-known in
probability theory~\cite{KingmanTaylor} 
that this alone implies that $n$-th order cumulants (or connected n-point 
correlation functions) of variables that are normalized sums of these independent source contributions 
scale in a characteristic way $\propto 1/N^{n-1}$ with the number $N$ of sources. Different eccentricities $\epsilon_m\{n\}$ have been calculated for such a model to various orders in $1/N$ by Bhalerao and Ollitrault \cite{Bhalerao:2006tp} as well as Alver et al.\ \cite{Alver:2008zza} and where shown to quantitatively reproduce results of more sophisticated Glauber models in nucleus-nucleus collisions \cite{Bhalerao:2011bp}. 
First indications that the scaling with $n$ is particularly relevant for initial conditions in proton-nucleus and nucleus-nucleus
collisions go back to numerical findings of Bzdak, Bozek and McLerran \cite{Bzdak:2013rya}. They were sharpened subsequently due to work of Ollitrault and Yan \cite{Yan:2013laa} (see also Bzdak and Skokov \cite{Bzdak:2013raa}) who established a related scaling  for eccentricity cumulants
at vanishing impact parameter in an analytically accessible model of independent point sources (IPSM)
and who showed that this reproduces with good numerical accuracy  the eccentricity cumulants 
in other currently used models of initial conditions.  

The present paper aims at contributing to this important recent development. To this end, we shall
show how one can solve the IPSM completely, including the set of $n$-point correlation 
functions that characterize completely the information about the radial and azimuthal dependence
at zero and non-zero impact parameter. Based on this differential information, we shall provide
further evidence that the IPSM shares indeed important commonalities with realistic
model distributions. At finite impact parameter $b$, we shall find that the $1/N^{n-1}$-scaling
is broken for azimuthally averaged event samples. However, for small $b$, the leading $b$-dependence of 
the terms that break this scaling can be given analytically. Thus information about this $b$-dependence,
combined with information about the $1/N^{n-1}$-scaling for $b=0$ can provide an 
ordering principle that applies more generally to $n$-point correlators at zero and non-zero impact
parameter. We shall also discuss how the connected $n$-point correlation functions of initial fluctuations 
enter the calculation of measurable correlators of flow coefficients, and we shall point to 
possible further phenomenological applications of these insights. 

\begin{figure}
\centering
\includegraphics[width=0.55\textwidth]{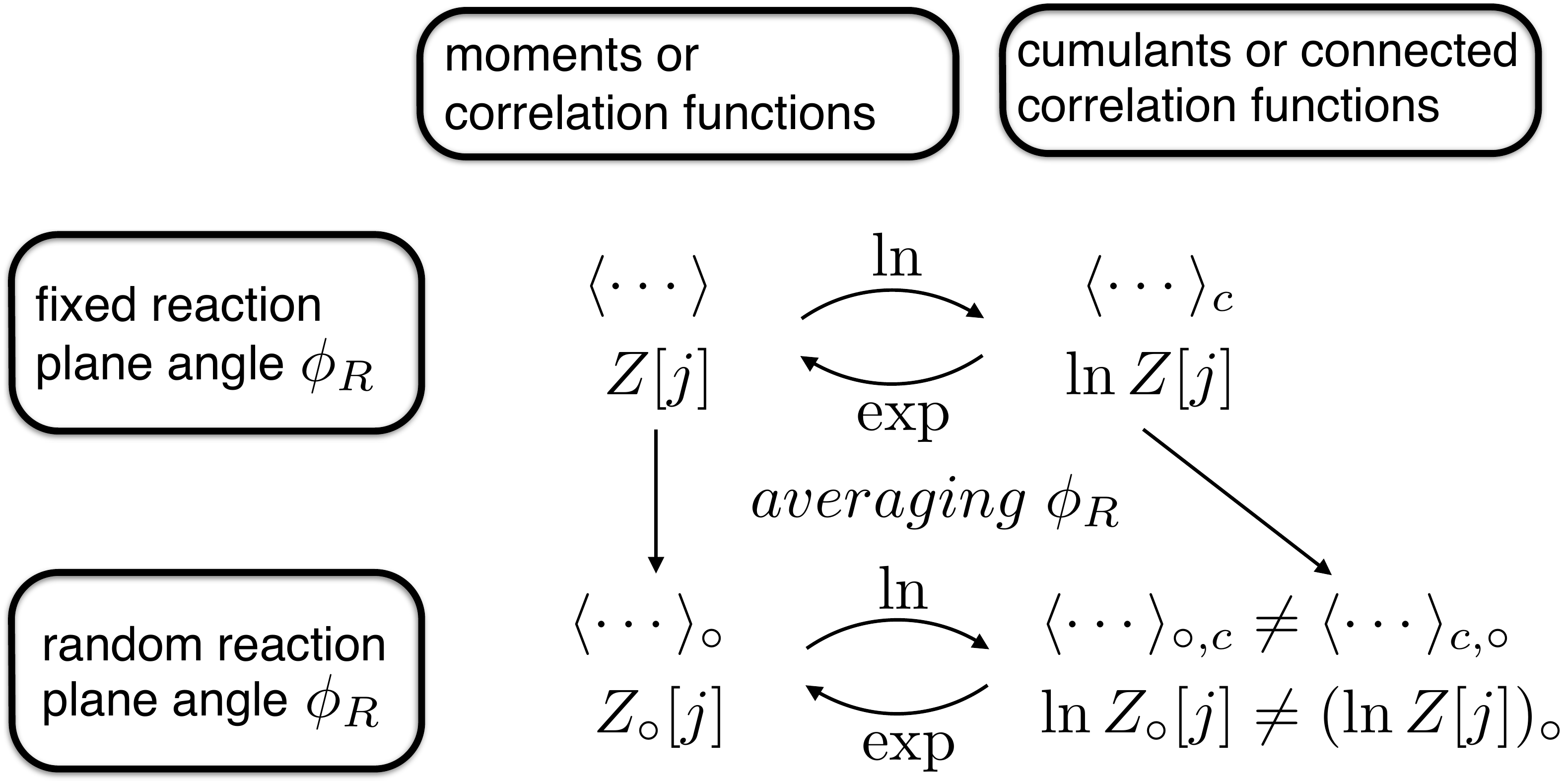}
\caption{Schematic overview of the operations with which we construct moments or correlation functions and the corresponding cumulants or connected correlation functions for event ensembles with fixed and with random reaction plane angle $\phi_R$. The operations of averaging over $\phi_R$ and of forming connected correlation functions do not commute, see text for further details.}
\label{fig0}
\end{figure}

The main assumption underlying the scaling of connected $n$-point correlation functions with $1/N^{n-1}$ is that the transverse density is given by a sum of $N$ independent and identically distributed random variables or functions of random variables.\footnote{In the concrete realization of the IPSM, these random variables are positions of point-like sources but the scaling with $N$  actually holds also for extended sources. The azimuthal and radial dependences of correlation functions change in that case, however. On the other side, the $1/N^{n-1}$ scaling gets violated as soon as correlation effects between the random variables such as e.\ g.\ excluded volume or other interaction effects are taken into account.}
As we discuss in more detail in the main text, this holds also for ensembles of non-central events, 
however, only if impact parameter and reaction 
plane orientation are kept fixed. In contrast, the phenomenologically relevant connected $n$-point
correlation functions are defined for ensembles with random azimuthal orientation. 
To cope with this complication, we find it useful to work in a framework sketched in Fig.~\ref{fig0}: 
we denote event averages with fixed azimuthal orientation by $\langle \ldots \rangle$ and we
construct moments and the corresponding cumulants as usual from a generating functional and its
logarithm, respectively. Randomizing the azimuthal orientation $\phi_R$ in the averages 
$\langle \ldots \rangle$ defines the average $\langle \ldots \rangle_\circ$. The scaling with 
$1/N^{n-1}$ is broken for the ensemble average $\langle \ldots \rangle_\circ$ 
at finite impact parameter, since the operation of averaging over $\phi_R$ does not commute
with the operation of passing from moments to cumulants. In other words, the cumulants with
respect to the randomized ensemble do not correspond to $\phi_R$-averages of cumulants 
evaluated at fixed $\phi_R$.  

As a significant part of this paper will study in detail the independent point-sources model, we conclude
this introduction by asking to what extent the spatial dependence of correlation functions in the IPSM can be expected to have physical significance. One may argue that the long-wavelength excitations (small values of azimuthal wave numbers $m$ and radial wave numbers $l$ in a Bessel-Fourier expansion) do not resolve the differences between a spatially extended but short range source function and a point-like source. Since these long wavelength modes are most important for the fluid dynamic evolution (others get damped quickly by dissipative effects), one might expect that also some space-dependent features of the independent point-sources model contain realistic aspects. They are universal in the sense that a larger class of models with extended sources (and even some early non-equilibrium dynamics as long as it is local) lead to equal correlation functions for the long wavelength modes. If it could be established, such a universality for the correlations of the most important fluid dynamic modes would have profound consequences. For instance,
in a mode-by-mode fluid dynamics framework one could use this knowledge of initial conditions for a detailed comparison between experimental results on correlations of harmonic flow coefficients and fluid dynamic calculations which would allow for a more detailed  determination of thermodynamic and transport properties.
These are some of the considerations that have prompted the following analysis.

\section{Flow cumulants}
\label{sec2}
In this section, we discuss how flow measurements are related to the $n$-mode correlation functions
of initial density perturbations that we are going to analyze in sections~\ref{sec3} and ~\ref{sec4} below. 
To focus on the structure of this relation, we shall defer some technical definitions to section~\ref{sec3}.
 We start from a perturbative expansion of the complex-valued event-wise flow coefficients in powers
 of weights $w_l^{(m)}$ that characterize these density perturbations in terms of 
 azimuthal ($m$) and radial ($l$) wave numbers~\cite{Floerchinger:2013tya}
\begin{eqnarray}
	V_{m}^* \equiv v_m e^{-i\, m\, \psi_m} &=& \sum_{m_1,l_1} S_{(m_1) l_1} \, w^{(m_1)}_{l_1} \, \delta_{m,m_1}
	\nonumber \\
	&& + \sum_{m_1,m_2,l_1,l_2} S_{(m_1,m_2) l_1,l_2} \, w^{(m_1)}_{l_1}\, w^{(m_2)}_{l_2}  \, \delta_{m,m_1+m_2}
	\nonumber \\
	&&+ \sum_{m_1,m_2,m_3,l_1,l_2,l_3} S_{(m_1,m_2,m_3) l_1,l_2,l_3} \, w^{(m_1)}_{l_1}\, w^{(m_2)}_{l_2}\, w^{(m_3)}_{l_3} \, \delta_{m,m_1+m_2+m_3} 
	\nonumber \\
	&&+ \ldots
	\label{eq2.1}
\end{eqnarray} 
Here, the indices $m_i$ are summed over the range $(-\infty\ldots \infty)$ and the indices $l_i$ are summed over the range $(1,\ldots,\infty)$. As coefficients of a Bessel-Fourier expansion, defined in eq.~(\ref{eq3.4}) below, the $w^{(m)}_l$ satisfy $w^{(m)}_l=(-1)^m w^{(-m)*}_l$. The dynamical response functions  $S_{(m_1,\ldots,m_n) l_1,\ldots,l_n}$ satisfy then $S_{(m_1,\ldots,m_n) l_1,\ldots,l_n} = (-1)^{m_1+\ldots+m_n} S^*_{(-m_1,\ldots,-m_n)l_1,\ldots,l_n}$. For the harmonic flow coefficients one has $V_{-m} = V_{m}^*$.
In general, $n$-th order flow cumulants $v_m\lbrace n\rbrace^n$ denote the connected $n$-point event average of flow coefficients $V_m$. The lowest order cumulants take the explicit form~\cite{Borghini:2000sa,Borghini:2001vi}
\begin{eqnarray}
	v_m\lbrace 2 \rbrace^2 &\equiv& \left\langle V_mV_{-m} \right\rangle_\circ\, ,
		\label{eq2.2} \\
		v_{m}\lbrace 4 \rbrace^4 &\equiv& - \langle (V_m V_{-m})^2\rangle_\circ	
			+ 2\, \langle  V_m V_{-m} \rangle_\circ^2\, ,
			\label{eq2.3} \\
	v_m\lbrace 6 \rbrace^6 &\equiv& \frac{1}{4} \left[ \langle \left(V_mV_{-m}\right)^3 \rangle_\circ	
		- 9\, \langle \left(V_mV_{-m}\right)^2 \rangle_\circ\, \langle V_mV_{-m}\rangle_\circ
			+ 12\, \langle V_mV_{-m} \rangle_\circ^3 \right]	\, . 
			\label{eq2.4}	
\end{eqnarray}
These higher order flow cumulants are measured in ion-ion and in proton-ion collisions \cite{ALICE:2011ab, Aad:2013fja, Chatrchyan:2013nka}. 
With the help of the perturbative expansion (\ref{eq2.1}), one can write flow cumulants 
as products of event averages of initial fluctuating modes $w_l^{(m)}$ times dynamical response functions $S_{(m_1,\ldots,m_n) l_1,\ldots,l_n}$. For the second 
order flow cumulant, one finds up to fifth order in initial fluctuations
\begin{eqnarray}
	v_m\lbrace 2 \rbrace^2 
			=  \langle T_2 \rangle_\circ 	+  \langle T_{3A} \rangle_\circ+  \langle T_{3B}\rangle_\circ +  \langle T_{4A}\rangle_\circ +  \langle T_{4B}\rangle_\circ+  \langle T_{4C}\rangle_\circ + {\cal O}\left(w^5\right)\, ,
		\label{eq2.5}
\end{eqnarray}
where
\begin{eqnarray}
		T_{2} &=& \sum_{l_1,l_2} S_{(m)l_1} \, S_{(-m)l_2} \;
		w^{(m)}_{l_1} w^{(-m)}_{l_2}  \, ,
		 \nonumber \\
		T_{3A} &=& \sum_{\substack{m_2,m_3,\\l_1,l_2,l_3}}  S_{(m) l_1}
		 S_{(-m_2, -m_3) l_2,l_3} \,
		 w^{(m)}_{l_1} 
		 w^{(-m_2)}_{l_2}
		 w^{(-m_3)}_{l_3} \, \delta_{m,m_2+ m_3}\, ,
				\nonumber \\
		T_{3B} &=&  \sum_{\substack{m_1,m_2,\\l_1,l_2,l_3}} S_{(m_1,m_2) l_1,l_2}
		S_{(-m) l_3} \,
		w^{(m_1)}_{l_1}
		w^{(m_2)}_{l_2}
		w^{(-m)}_{l_3} \delta_{m,m_1+m_2}\, ,
		\nonumber \\
		T_{4A} &=& \sum_{\substack{m_1,\ldots,m_4,\\l_1,\ldots,l_4}} \hspace*{-0.2cm} S_{(m_1,m_2)l_1,l_2} 
			S_{(-m_3, -m_4)l_3,l_4}\, 
		w^{(m_1)}_{l_1}\, 
	         w^{(m_2)}_{l_2}\,
		 w^{(-m_3)}_{l_3}\, 
		w^{(-m_4)}_{l_4} \delta_{m, m_1+ m_2}\, \delta_{m,m_3+m_4}\, ,
		\nonumber \\
		T_{4B} &=&  \sum_{\substack{m_2,m_3,m_4,\\l_1,l_2,l_3,l_4}}		
		S_{(m) l_1}
		S_{(-m_2, -m_3, -m_4) l_2,l_3,l_4} \,  
		w^{(m)}_{l_1} 
		w^{(-m_2)}_{l_2}\, 
		w^{(-m_3)}_{l_3}\, 
		w^{(-m_4)}_{l_4}\, \delta_{m,m_2+m_3+m_4}\, ,
		 \nonumber \\
		T_{4C} &=&  \sum_{\substack{m_1,m_2,m_3,\\l_1,l_2,l_3,l_4}}
		S_{(m_1,m_2,m_3)l_1,l_2,l_3}
		         S_{(-m)l_4}\, 
			w^{(m_1)}_{l_1}\, 
			w^{(m_2)}_{l_2}\, 
			w^{(m_3)}_{l_3} \, 
		         w^{(-m)}_{l_4}\, \delta_{m,m_1+m_2+m_3}\, .
		\label{eq2.6}
\end{eqnarray}
Here, $\langle T_2 \rangle_\circ$ is the only term that involves only the linear dynamic response terms $S_{(m)l}$ of the perturbative series (\ref{eq2.1}). It is thus the entire linear response contribution to $v_m\lbrace 2 \rbrace^2$.\footnote{The parametrization of initial fluctuations 
in terms of eccentricities would amount to neglecting radial wave numbers and substituting $w^{(m)}_l \rightarrow \epsilon_m$ in our discussion. The linear response contribution to (\ref{eq2.5}) reduces then to the well-known approximate linear relation $v_m\lbrace 2 \rbrace^2 \propto \epsilon_m\lbrace 2\rbrace^2$, which is at the basis of participant eccentricity scaling~\cite{Teaney:2010vd,Niemi:2012aj}. }
The terms $ \langle T_{3A} \rangle_\circ$, $ \langle T_{3B} \rangle_\circ$ ... are higher order (non linear) corrections to this linear response. In eq.\ \eqref{eq2.6} we have included terms up to order ${\cal O}(w^5)$ to display the first non-vanishing correction to linear dynamics in a Gaussian model of initial conditions where it arises at order $w^4$ (see section \ref{sec3}).

In the same way, we can write the fourth order flow up to seventh order in initial fluctuations,
\begin{eqnarray}
v_{m}\lbrace 4 \rbrace^4 
			&=& - \langle T_2\, T_2 \rangle_\circ + 2 \langle T_2\rangle_\circ^2
		        \nonumber \\
			&& 
			- 2 \langle T_2 \left( T_{3A} + T_{3B} \right) \rangle_\circ 
			+ 4 \langle T_2 \rangle_\circ\, \langle  T_{3A} + T_{3B} \rangle_\circ 
			\nonumber \\
			&& - \langle \left(T_{3A} + T_{3B} \right) \left( T_{3A} + T_{3B} \right) \rangle_\circ
				+ 2  \langle \left(T_{3A} + T_{3B} \right) \rangle_\circ^2
		\nonumber \\
		&& - 2\, \langle T_2\, T_{4A} \rangle_\circ  + 4\, \langle T_2\rangle_\circ \, \langle T_{4A} \rangle_\circ +
	{\cal O}({w}^7)\, .
		\label{eq2.7}
\end{eqnarray}
The linear response term of (\ref{eq2.7}) 
can be written in terms of a connected four-point function of initial fluctuations~\footnote{
This is well-known, of course. For a parametrization of initial fluctuations 
in terms of eccentricities ($w^{(m)}_l \rightarrow \epsilon_m$),  the linear response contribution
to (\ref{eq2.7}) would reduce to the well-known approximate ansatz $v_m\lbrace 4 \rbrace^2 \propto \epsilon_m\lbrace 4\rbrace^2$ in terms
of a connected 4-point function of initial eccentricities $\epsilon_m$.},
\begin{equation}
	- \langle T_2\, T_2 \rangle_\circ + 2 \langle T_2\rangle_\circ^2 = - S_{(m)l_1}\, S_{(-m)l_2}\, S_{(m)l_3} \, S_{(-m)l_4}\,  
	\langle w^{(m)}_{l_1} w^{(-m)}_{l_2} w^{(m)}_{l_3} w^{(-m)}_{l_4} \rangle_{\circ,c}\, .
%	C^{(m,m,m,m)}_{l_1,\bar{l}_1,l_2,\bar{l}_2}\, .
	\label{eq2.8}
\end{equation}
(Summation over the indices $l_1,\ldots,l_4$ is implied here and in the following.)
In general, the linear response contribution to $v_m\lbrace{ 2n\rbrace}^{2n}$ is proportional to a
connected $(2n)$-mode correlator
\begin{equation}
	v_m\lbrace{ 2n\rbrace}^{2n} = \langle \prod_{i=1}^n \left( S_{(m)l_i}\, S_{(-m)l^\prime_i}
		w^{(m)}_{l_i}\, w^{(-m)}_{l^\prime_i} \right) \rangle_{\circ,c}
		+ \hbox{non-lin. dynamic response} \, .
		\label{eq2.9}
\end{equation}
Flow measurements are not limited to the determination of flow cumulants. In principle, arbitrary 
event averages $\langle V_{m_1}\, V_{m_2}\, \ldots V_{m_n}\,\rangle_\circ$ of products of flow coefficients 
are experimentally accessible, see e.\ g.\ \cite{Aad:2014fla}. For event samples with randomized orientation of the reaction plane, the
simplest generalization are 3-flow correlators $\langle V_{m_1}\, V_{m_2}\, V_{m_3}\,\rangle_\circ$
with $\sum_{i=1}^3 m_i = 0$. To be specific, let us write here the expansion of one of them,
\begin{eqnarray}
	\langle V_{2}\, V_{3}\, V_{5}^*\,\rangle_\circ &=& S_{(2)l_2} S_{(3)l_3} S_{(-5)l_5}
		\langle w^{(2)}_{l_2}\, w^{(3)}_{l_3}\, w^{(-5)}_{l_5} \,\rangle_\circ \nonumber \\
		&& + S_{(2)l_2} S_{(3)l_3} S_{(-2,-3)l_5,\bar l_5}
		\langle w^{(2)}_{l_2}\, w^{(3)}_{l_3}\, w^{(-2)}_{l_5}\, w^{(-3)}_{\bar l_5} \,\rangle_\circ \nonumber \\
		&& + S_{(2)l_2} S_{(5,-2)l_3,\bar l_3} S_{(-5)l_5} 
		\langle w^{(2)}_{l_2}\, w^{(5)}_{l_3}\, w^{(-2)}_{\bar l_3}\, w^{(-5)}_{l_5} \,\rangle_\circ \nonumber \\
		&& + S_{(5,-3)l_2,\bar l_2} S_{(3)l_3} S_{(5)l_5}
		\langle w^{(5)}_{l_2}\, w^{(-3)}_{\bar l_2}\, w^{(3)}_{l_3}\, w^{(-5)}_{l_5} \,\rangle_\circ 
		+ \ldots\, . \label{eq2.10}
\end{eqnarray}
On the right hand side we have included terms from linear dynamics as well as those quadratic corrections that contain four point functions with two opposite index pairs $(m,-m)$. These are the leading contributions for ensembles that are close to Gaussian. 

Other experimentally easily accessible 3-flow correlators include 
$\langle V_{2}\, V_{2}\, V_{4}^*\,\rangle_\circ$ and $\langle V_{3}\, V_{3}\linebreak[3] \, V_{6}^*\,\rangle_\circ$, for 
which similar expansions can be written down. The dynamical response functions that appear 
on the right hand side of (\ref{eq2.10}) can be found also in the expansion of the flow cumulants 
(\ref{eq2.9}).~\footnote{This is obvious for the linear response terms $S_{(2)l_2}$, $S_{(3)l_3}$ and 
$S_{(5)l_5}$, but one can check for instance easily that the non-linear response terms 
in (\ref{eq2.10}) appear in the contributions $T_{3A}$ and $T_{3B}$ that enter the flow cumulants 
(\ref{eq2.5}) and (\ref{eq2.7}). In particular, $S_{(2,3)l_5,\bar l_5}$ appears as a non-linear contribution
to $v_5\lbrace 2 \rbrace^2$ and $v_5\lbrace 4 \rbrace^4$. } But in the 3-flow correlators, the linear
and non-linear dynamic response terms are weighted with a different set of informations about the 
initial conditions, namely a different set of moments 
$\langle w_{l_1}^{(m_1)} \ldots w_{l_n}^{(m_n)} \rangle_\circ$ that typically involve harmonic modes with different $m$. 

As illustrated by the examples discussed so far, the calculation of flow correlation measurements 
$\langle V_{m_1}\, V_{m_2}\, V_{m_3} \ldots \,\rangle_\circ$ requires knowing the 
initial n-mode correlators $\langle w_{l_1}^{(m_1)} \ldots \linebreak[3]  w_{l_n}^{(m_n)} \rangle_\circ$ and the
dynamical response functions $S_{(m_1,\ldots,m_n) l_1,\ldots,l_n}$. We note that the 
dynamical response functions are known {\it in principle}, in the sense that they are calculable 
once the thermodynamic information entering hydrodynamic evolution and the event-averaged
initial enthalpy density is given. No further model dependent assumption enters their calculation.
A method of how to determine them numerically was given in Ref.~\cite{Floerchinger:2013tya}. 
On the other hand, the correlators $\langle w_{l_1}^{(m_1)} \ldots w_{l_n}^{(m_n)} \rangle_\circ$ 
should be calculable {\it in principle} from a microscopic theory of thermalization dynamics. 
In practice, however, this program is not yet carried out, and the
initial conditions are currently regarded as the most significant source of uncertainties in the 
calculation of flow observables. This motivates us to investigate in the following what can be said
on the basis of general considerations about the structure of $n$-mode correlators 
$\langle w_{l_1}^{(m_1)} \ldots w_{l_n}^{(m_n)} \rangle_\circ$.

%%%%%%%%%%%%%%%%%%%%%%%%%%%%%%%%%%%%%%%%%%%%%%%%%%%%%%%%%%%%%%%%%%
\section{Gaussian probability distributions of initial conditions}
\label{sec3}

In this section, we introduce Gaussian probability distributions of  fluctuations in the initial transverse enthalpy density $w(\vec x)$,  
and we discuss their implications for flow cumulants and flow probability distributions. 

%%%%%%%%%%%%%%%%%%%%%%%%%%%%%%%%%%%
\subsection{Gaussian model of initial fluctuations for fixed reaction plane angle $\phi_R$}
\label{sec3a}
We start from the general form of a Gaussian probability distribution of the enthalpy density written for fixed impact parameter and reaction 
plane angle $\phi_R$ (see also appendix C of ref.\ \cite{Floerchinger:2013vua})
\begin{equation}
p[w] = {\cal N} \exp \left( -\frac{1}{2}\int d^2x d^2 y \left[ w(\vec x) - \bar w(\vec x)\right] M(\vec x, \vec y) \left[w(\vec y) -\bar w(\vec y) \right] \right).
\label{eq3.1}
\end{equation}
As a Gaussian distribution it is specified completely in terms of the expectation value $\bar w(\vec x)$ and the connected two-point correlation function $C(\vec x, \vec y)$, which is the inverse of $M(\vec x, \vec y)$ seen as a matrix of infinite dimension with indices $\vec x$ and $\vec y$. For an arbitrary event, we write the enthalpy density in a Bessel-Fourier expansion
\begin{equation}
w(r,\phi) = w_\text{BG}(r) \left[ 1 + \sum_{m=-\infty}^\infty \sum_{l=1}^\infty w^{(m)}_l e^{im\phi} J_m\left(z^{(m)}_l \rho(r)\right) \right].
\label{eq3.4}
\end{equation}
Here, $\rho(r)$ is a monotonous function that maps $r\in (0,\infty)$ to $\rho\in(0,1)$. 
It is specified in appendix \ref{appb}. The real numbers
$z^{(m)}_l$ denote the $l$'th zeroes of the Bessel function $J_m(z)$.   The Bessel-Fourier coefficients $w^{(m)}_l$ are complex (the phase contains 
information about the azimuthal orientation), but since the enthalpy density is real, the coefficients
satisfy $w^{(m)}_l = (-1)^m w^{(-m)*}_l$. An inverse relation that expresses $w^{(m)}_l$ in terms of 
$w(r,\phi)$ is given in Eq.\ \eqref{eq4.11}. 

The expectation value at fixed impact parameter and reaction plane angle $\phi_R$ can be written in the same Bessel-Fourier expansion,
\begin{equation}
\bar w(r,\phi) = w_\text{BG}(r) \left[ 1+\sum_{\substack{m=-\infty \\ m\; \text{even}}}^\infty \sum_{l=1}^\infty \bar w^{(m)}_l  e^{i m(\phi-\phi_R)} \; J_m\left( z^{(m)}_l \rho(r) \right) \right]\, .
\label{eq3.2}
\end{equation}
Here, the sum over $m$ on the right hand side goes only over the even values $m=\pm 2, \pm 4, \ldots$ as it follows from the discrete symmetry that $\bar w(r,\phi) = \bar w(r,\phi+\pi)$. The function $w_\text{BG}(r)$ is defined such that the $m=0$ component in the sum vanishes. The dimensionless and real coefficients $\bar w^{(m)}_l$ depend on centrality and they vanish with vanishing impact parameter 
$b$, i.e., for ultra-central collisions. One can show that for small $b$ they behave like $w^{(m)}_l \sim b^{|m|}$, see appendix~\ref{appd}. For the two-point correlation function we write a Bessel-Fourier expansion in terms of the coefficients $C^{(m_1,m_2)}_{l_1,l_2}$,
\begin{equation}
\begin{split}
C(r_1,r_2,\phi_1, \phi_2) = w_\text{BG}(r_1) \, w_\text{BG}(r_2) & \sum_{m_1,m_2 = -\infty}^\infty \sum_{l_1,l_2 = 1}^\infty C^{(m_1,m_2)}_{l_1,l_2} \, e^{im_1(\phi_1-\phi_R)} \, e^{im_2(\phi_2-\phi_R)} \\
& \times J_{m_1}\left( z^{(m_1)}_{l_1} \rho(r_1) \right) \, J_{m_2}\left( z^{(m_2)}_{l_2} \rho(r_2) \right).
\end{split}
\label{eq3.3}
\end{equation}
Since $C(r_1,r_2,\phi_1,\phi_2)$ is real one has $C^{(m_1,m_2)}_{l_1,l_2} = (-1)^{m_1+m_2} C^{(-m_1,-m_2)*}_{l_1,l_2}$.
Note that Eq.\ \eqref{eq3.3} contains a factor $e^{-i(m_1+m_2)\phi_R}$ such that the right hand side vanishes when averaged over the reaction plane angle $\phi_R$ with uniform distribution, except for $m_1+m_2=0$.
The expectation value calculated for an event sample with fixed orientation of the reaction plane
reads now
\begin{equation}
\langle w^{(m)}_l \rangle = \bar w^{(m)}_l e^{-im \phi_R}\, ,
\label{eq3.5}
\end{equation}
and the two-mode correlation function is
\begin{equation}
\begin{split}
\langle w^{(m_1)}_{l_1} w^{(m_2)}_{l_2} \rangle = & \left[ C^{(m_1,m_2)}_{l_1,l_2} + \bar w^{(m_1)}_{l_1} \bar w^{(m_2)}_{l_2}  \right]  e^{-i(m_1+m_2) \phi_R}\, .
\end{split}
\label{eq3.6}
\end{equation}
The probability distribution in Eq.\ \eqref{eq3.1} can then be written as a function of the (complex) Bessel-Fourier coefficients $w^{(m)}_l$,
\begin{equation}
\begin{split}
p[w] = {\cal N}\exp {\Bigg(} -\frac{1}{2} \sum_{m_1,m_2,l_1,l_2} & \left[ w^{(m_1)}_{l_1} - \bar w^{(m_1)}_{l_1} e^{-im_1 \phi_R} \right]\\
\times & \left[ w^{(m_2)}_{l_2} - \bar w^{(m_2)}_{l_2} e^{-im_2 \phi_R} \right] \; T^{(m_1,m_2)}_{l_1,l_2} e^{i(m_1+m_2)\phi_R} {\Bigg)},
\label{eq3.7}
\end{split}
\end{equation}
where $T^{(m_1,m_2)}_{l_1,l_2}$ is the inverse of $C^{(m_1,m_2)}_{l_1,l_2}$ as a matrix with indices $(m_1,l_1)$ and $(m_2,l_2)$. Higher $n$-mode correlation functions can be calculated directly from $p[w]$, but it is convenient to derive them as $n$-th derivatives with respect to the source terms of the partition function 
\begin{eqnarray}
Z[j] &=& \left\langle \exp\left( \sum_{m=-\infty}^\infty \sum_{l=1}^\infty j^{(-m)}_l w^{(m)}_l \right) \right\rangle.
\nonumber \\
% \begin{split}
&=& \exp{\Bigg(}  \sum_{m=-\infty}^\infty \sum_{l=1}^\infty j^{(-m)}_l \bar w^{(m)}_l e^{-im \phi_R} 
\nonumber \\
&& \qquad + \frac{1}{2} \sum_{m_1,m_2 = - \infty}^\infty \sum_{l_1,l_2=1}^\infty j^{(-m_1)}_{l_1} j^{(-m_2)}_{l_2} C^{(m_1,m_2)}_{l_1,l_2} e^{-i(m_1+m_2)\phi_R} 
 {\Bigg )}.
% \end{split}
\label{eq3.8}
\end{eqnarray}
This equation shows nicely that the Gaussian model for a particular centrality class needs as an input besides the background density $w_\text{BG}(r)$ only the expectation values $\bar w^{(m)}_l$ that can be determined from geometrical considerations, and the two-point correlator $C^{(m_1,m_2)}_{l_1,l_2}$. In particular, the three-mode
correlator takes the form
\begin{equation}
\begin{split}
\langle w^{(m_1)}_{l_1} w^{(m_2)}_{l_2} w^{(m_3)}_{l_3} \rangle = & {\Big [} C^{(m_1,m_2)}_{l_1,l_2} \bar w^{(m_3)}_{l_3}
+ C^{(m_2,m_3)}_{l_2,l_3} \bar w^{(m_1)}_{l_1}
+ C^{(m_3,m_1)}_{l_3,l_1} \bar w^{(m_2)}_{l_2} \\
& + \bar w^{(m_1)}_{l_1} \bar w^{(m_2)}_{l_2} \bar w^{(m_3)}_{l_3}{\Big ]} \; e^{-i(m_1+m_2+m_3) \phi_R}\, ,
\end{split}
\label{eq3.9}
\end{equation}
and the four-point correlation function reads
\begin{equation}
\begin{split}
\langle w^{(m_1)}_{l_1} w^{(m2)}_{l_2} w^{(m_3)}_{l_3} & w^{(m_4)}_{l_4} \rangle = 
{\Big [} C^{(m_1,m_2)}_{l_1,l_2} C^{(m_3,m_4)}_{l_3,l_4} 
+ C^{(m_1,m_3)}_{l_1,l_3} C^{(m_2,m_4)}_{l_2,l_4}  
+ C^{(m_1,m_4)}_{l_1,l_4} C^{(m_2,m_3)}_{l_2,l_3} \\
& + C^{(m_1,m_2)}_{l_1,l_2} \bar w^{(m_3)}_{l_3} \bar w^{(m_4)}_{l_4}
+ C^{(m_1,m_3)}_{l_1,l_3} \bar w^{(m_2)}_{l_2} \bar w^{(m_4)}_{l_4}
+ C^{(m_1,m_4)}_{l_1,l_4} \bar w^{(m_2)}_{l_2} \bar w^{(m_3)}_{l_3}\\
& + C^{(m_2,m_3)}_{l_2,l_3} \bar w^{(m_1)}_{l_1} \bar w^{(m_4)}_{l_4}
+ C^{(m_2,m_4)}_{l_2,l_4} \bar w^{(m_1)}_{l_1} \bar w^{(m_3)}_{l_3}
+ C^{(m_3,m_4)}_{l_3,l_4} \bar w^{(m_1)}_{l_1} \bar w^{(m_2)}_{l_2} \\ 
& + \bar w^{(m_1)}_{l_1} \bar w^{(m_2)}_{l_2} \bar w^{(m_3)}_{l_3} \bar w^{(m_4)}_{l_4} {\Big ]} \; e^{-i(m_1+m_2+m_3+m_4) \phi_R}.
\end{split}
\label{eq3.10}
\end{equation}
The connected correlation functions can be obtained from derivatives of $\ln Z[j]$. The connected two-mode correlator equals the connected part of eq.~(\ref{eq3.6}), and the connected correlators of more 
than two modes vanish of course for this Gaussian distribution.

%%%%%%%%%%%%%%%%%%%%%%%%%%%%%%%%%%%%%%%%%%%%%%%%%%%%%%%%%%%%%%%%
\subsection{Averaging the Gaussian model of initial fluctuations over $\phi_R$}
\label{sec3b}
So far, we have discussed event averages for ensembles with fixed reaction plane $\phi_R$. However, 
essentially all measurements are for ensembles with randomized orientation of the reaction plane. 
One can formally introduce a distribution for an ensemble of events with random orientation by 
averaging over $\phi_R$,
\begin{equation}
p_\circ[w] = \frac{1}{2\pi} \int_0^{2\pi} d \phi_R \; p[w].
\label{eq3.11}
\end{equation}
Event averages evaluated with this azimuthally symmetric probability distribution will be denoted in the following by $\langle \dots \rangle_\circ$. It is then a consequence of azimuthal symmetry that
\begin{eqnarray}
	\langle w^{(m)}_l \rangle_\circ &=&0\, ,
	\nonumber \\
	 \langle w^{(m_1)}_{l_1} w^{(m_2)}_{l_2} \rangle_\circ &=& 
	\left[ C^{(m_1,m_2)}_{l_1,l_2} + \bar w^{(m_1)}_{l_1} \bar w^{(m_2)}_{l_2}  \right]  \delta_{m_1,-m_2}\, .
	\label{eq3.12}
\end{eqnarray}
Similarly, the 3-mode and 4-mode correlation functions for an ensemble of randomized azimuthal orientation can be obtained from equations 
(\ref{eq3.9}) and (\ref{eq3.10}) by averaging over $\phi_R$. The result is obtained from (\ref{eq3.9}) and 
(\ref{eq3.10}) by replacing on the left hand side of these equations $\langle \dots \rangle$ with $\langle \dots \rangle_\circ$ 
and by replacing on the right hand side the phases $\exp\left(- {i (\sum_j m_j)\phi_R}\right)$ by their $\phi_R$-integrals which are Kronecker-$\delta$'s,
$\delta_{\sum_j m_j, 0}$.

It is important to note that, in general, $p_\circ[w]$ in (\ref{eq3.11}) is not a Gaussian distribution even if $p[w]$ is one. As a consequence,
the connected higher-mode correlators do not vanish for the azimuthally randomized average 
$\langle \dots \rangle_\circ$. To illustrate this point further, we write the connected 4-mode correlator 
that appears in the linear response term to $v_m\lbrace 4 \rbrace^4$, 
\begin{eqnarray}
\langle w^{(m)}_{l_1} w^{(-m)}_{l_2} w^{(m)}_{l_3} w^{(-m)}_{l_4} \rangle_{\circ,c} &=& 
\langle w^{(m)}_{l_1} w^{(-m)}_{l_2} w^{(m)}_{l_3} w^{(-m)}_{l_4} \rangle_\circ
\nonumber \\
&& - \langle w^{(m)}_{l_1} w^{(-m)}_{l_2}\rangle_\circ    \langle w^{(m)}_{l_3}  w^{(-m)}_{l_4} \rangle_\circ
\nonumber \\
&& 
- \langle w^{(m)}_{l_1} w^{(-m)}_{l_4}\rangle_\circ    \langle w^{(m)}_{l_3}  w^{(-m)}_{l_2} \rangle_\circ
\nonumber \\
&=& 
 C^{(m,m)}_{l_1,l_3} C^{(-m,-m)}_{l_2,l_4} 
+ C^{(m,m)}_{l_1,l_3} \bar w^{(-m)}_{l_2} \bar w^{(-m)}_{l_4}
\nonumber \\
&& + C^{(-m,-m)}_{l_2,l_4} \bar w^{(m)}_{l_1} \bar w^{(m)}_{l_3}
 - \bar w^{(m)}_{l_1} \bar w^{(-m)}_{l_2} \bar w^{(m)}_{l_3} \bar w^{(-m)}_{l_4}  \, .
\label{eq3.13}
\end{eqnarray}
Here, $C^{(m,m)}_{l_1,l_3}$ and $C^{(-m,-m)}_{l_2,l_4} $ are defined as the connected two-mode correlators  with respect to the event average for fixed $\phi_R$,
see equation~(\ref{eq3.6}), while the corresponding components of the connected two-mode correlator for an azimuthally randomized event average vanish, see (\ref{eq3.12}). We can now make the following remarks about general properties
of the probability distribution $p_\circ[w]$:
\begin{enumerate}
	\item For {\it vanishing impact parameter}, the probability distribution (\ref{eq3.7}) becomes azimuthally 	symmetric even without averaging over $\phi_R$. This implies
	\begin{equation}
		\bar w_l^{(m)} = 0\qquad \hbox{for $b=0$}.
		\label{eq3.14}
	\end{equation}
	Also, azimuthal symmetry of the event-averaged geometry implies that the two-point correlation 
	function (\ref{eq3.3}) can depend only on $\phi_1 - \phi_2$, and hence
	\begin{equation}
		C_{l_1,l_2}^{(m_1,m_2)} = C_{l_1,l_2}^{(m_1)}\, \delta_{m_1,-m_2}\qquad \hbox{for $b=0$}.
		\label{eq3.15}
	\end{equation}
	As a consequence, the probability distribution $p_\circ[w]$ is Gaussian in this limit, and all 
	connected higher-mode correlators vanish. The distribution is fully characterized by 
	$\langle w_{l_1}^{(m_1)} \, w_{l_2}^{(m_2)} \rangle_\circ = C_{l_1,l_2}^{(m_1)}\, \delta_{m_1,-m_2}$.
	\item At {\it finite impact parameter}, the two-point correlation function (\ref{eq3.3}) of fluctuations
	will depend in general not only on $\phi_1 - \phi_2$, but also on $\phi_1 + \phi_2$. 
	Event-averages are then still symmetric under reflections on the reaction plane,
	$\phi_{1,2} - \phi_R \rightarrow  \phi_R - \phi_{1,2} $. Invariance of (\ref{eq3.3})
         under this reflection symmetry implies
	\begin{equation}
		C_{l_1,l_2}^{(m_1,m_2)} = C_{l_1,l_2}^{(-m_1,-m_2)}\, ,
		\label{eq3.16}
	\end{equation}
	and thus the coefficients $C_{l_1,l_2}^{(m_1,m_2)} $ must be real. Moreover, event averages
	at finite impact parameter are symmetric under the rotation
	$\phi_{1,2} \rightarrow \phi_{1,2} + \pi$. This rotation changes each term on the right hand side of
	(\ref{eq3.3}) by a phase $e^{i(m_1+m_2)\pi}$. Therefore, invariance under rotation by $\pi$ implies
	\begin{equation}
		C_{l_1,l_2}^{(m_1,m_2)} = 0\qquad \hbox{for $m_1+m_2$ odd}\, .
		\label{eq3.17}
	\end{equation}
	However, if all  $C_{l_1,l_2}^{(m_1,m_2)}$ with $m_1 \not= -m_2$ would vanish, then (\ref{eq3.3})
	would depend only on $\phi_1 - \phi_2$, but not $\phi_1 + \phi_2$. This is not the most generic
	case as fluctuations will depend in general on the orientation with respect to the reaction plane 
	and therefore they will depend on $\frac{1}{2}\left(\phi_1 + \phi_2\right) - \phi_R$. From this, 
	we conclude that
	\begin{equation}
		C_{l_1,l_2}^{(m_1,m_2)} \not= 0\qquad \hbox{for some $m_1+m_2$ even and non-zero}\, .
		\label{eq3.18}
	\end{equation}	
	In particular, the coefficients $C_{l_1,l_2}^{(m,m)}$ and $C_{l_1,l_2}^{(-m,-m)}$ in the connected
	four-mode correlator (\ref{eq3.13}) can be expected to take non-vanishing values at finite impact
	parameter. The model discussed in section~\ref{sec4} provides an example for which
	these non-vanishing terms can be calculated explicitly, see eq.~(\ref{eq4.19}).
	\item The {\it event distribution in eccentricity} $\epsilon_m$ can be calculated from the probability 
	distribution $p[w]$. The eccentricity is (up to a small correction to normalization) 
	linear in the (complex) Bessel-Fourier coefficients $w^{(m)}_l$ \cite{Floerchinger:2013vua},
	\begin{equation}
	{\cal E}_m^* = \epsilon_m e^{ -im \psi_m} = \sum_{l=1}^\infty 
	{\cal K}^{(m)}_{\,\,l} w^{(m)}_l\, .
	\label{eq3.19}
	\end{equation}
	(The ${\cal K}^{(m)}_l$ are real with ${\cal K}^{(m)}_l = (-1)^m{\cal K}^{(-m)}_l$.)
	Since this is a linear relation, eq.\ \eqref{eq3.7} implies that at fixed reaction plane angle $\phi_R$ 
	the ${\cal E}_m$ are Gaussian distributed in the complex plane. An azimuthally
	randomized distribution for $\epsilon_m$ is obtained by integration over $\phi_R$
	\begin{eqnarray}
	p(\epsilon_m) &=&  \epsilon_m \int d\phi_R 
	\int D w \; \delta^{(2)}\left({\cal E}^*_m - \sum\nolimits_{l=1}^\infty {\cal K}^{(m)}_{\,\,l} 
	w^{(m)}_l \right) \; p[w] 
	\nonumber \\
	&=& \frac{\epsilon_m}{\pi\sqrt{\tau_m^2-\tau_m^{\prime 2}}} \int_0^{2\pi} d \phi 
	\nonumber \\
	&& \times \exp\left(-\frac{\epsilon_m^2 [ \tau_m-\tau_m^\prime \cos(2\phi) ] +\bar \epsilon_m^2 [\tau_m-\tau_m^\prime] - 2 \epsilon_m \bar \epsilon_m \cos(\phi) [\tau_m-\tau_m^\prime]}{\tau_m^2-\tau_m^{\prime 2}}\right)\, .\nonumber\\
	\label{eq3.20}
	\end{eqnarray}
	Here, the expectation value of the eccentricity at fixed $\phi_R$ is
	\begin{equation}
	\langle {\cal E}^*_m \rangle = \bar \epsilon_m e^{-im\phi_R}= 
	\sum_{l=1}^\infty {\cal K}^{(m)}_{\,\,l} \bar w^{(m)}_l e^{-im\phi_R}\, ,
	\label{eq3.21}
	\end{equation}
	and the two variance parameters are
	\begin{eqnarray}
	\tau_m &=& \sum_{l_1,l_2=1}^\infty  
	{\cal K}^{(m)}_{\,\,l_1} {\cal K}^{(-m)}_{\,\,l_2} \; C^{(m,-m)}_{l_1,l_2},
	\label{eq3.22}\\
	\tau_m^\prime &=& \sum_{l_1,l_2=1}^\infty  
	{\cal K}^{(m)}_{\,\,l_1} {\cal K}^{(m)}_{\,\,l_2} \; C^{(m,m)}_{l_1,l_2}\, .
	\label{eq3.23}
	\end{eqnarray}
	This distribution is well defined for $0< |\tau_m^\prime| \leq \tau_m$.
	\item At finite impact parameter, the remaining reflection symmetries of the event-averaged enthalpy
	density imply that 
	\begin{equation}
	 \bar w^{(m)}_{l} = 0\, \qquad \hbox{for $m$ odd,} 
	 \label{eq3.24} 
	\end{equation}
	and therefore
	\begin{equation}
	 \bar \epsilon_{m} = 0\, \qquad \hbox{for $m$ odd.}  
	 \label{eq3.25}
	\end{equation}
	For odd $m=1,3,5,\ldots$, one can then perform the integral over $\phi$ in equation (\ref{eq3.20})
	and one finds for the distribution in eccentricities
	\begin{equation}
	p(\epsilon_m) = 
	\frac{2 \epsilon_m}{\sqrt{\tau_m^2-\tau_m^{\prime 2}}} I_0\left( \frac{\tau_m^\prime\epsilon_m^2}{\tau_m^2-	\tau_m^{\prime 2}} \right) \exp\left( -\frac{\tau_m\epsilon_m^2}{\tau_m^2-\tau_m^{\prime 2}} \right) \, \qquad \hbox{for $m$ odd}.
	\label{eq3.26}
	\end{equation}
	\item According to (\ref{eq3.18}), $C_{l_1,l_2}^{(m_1,m_2)} $ is generally non-vanishing for even and 
	non-vanishing $m_1+m_2$. It is nevertheless interesting to investigate the {\it simplifying ad hoc 
	assumption} that $C_{l_1,l_2}^{(m_1,m_2)} = C_{l_1,l_2}^{(m_1)}\, \delta_{m_1,-m_2}$ at finite impact 
	parameter. For the event distribution (\ref{eq3.20}) in eccentricity, this corresponds to the case 
	$\tau_m^\prime=0$.  The integral over $\phi_R$ can then be done analytically and one finds 
	\begin{equation}
	p_{\rm BG}(\epsilon_m) = \frac{2 \epsilon_m}{\tau_m} I_0
	\left( 2 \frac{\epsilon_m \bar \epsilon_m}{\tau_m} \right) 
	\exp\left( - 	\frac{\epsilon_m^2 + \bar \epsilon_m^2}{\tau_m} \right).
	\label{eq3.27}
 	\end{equation}
	This is the ``Bessel-Gaussian'' distribution proposed in ref.\ \cite{Voloshin:1994mz,Voloshin:2007pc} and used by ATLAS 
	to compare to distributions of event-by-event flow harmonics~\cite{Aad:2013xma}.
\item Finally, for small impact parameter $b$ one has $\tau_m^\prime \sim b^{2m}$ and $\bar \epsilon_m \sim b^m$, see appendix~\ref{appd}. For $b\to 0$ the distribution in eq.\ \eqref{eq3.20} approaches a Gaussian distribution,
\begin{equation}
p(\epsilon_m) = \frac{2 \epsilon_m}{\tau_m} \exp\left( - \frac{\epsilon_m^2}{\tau_m} \right).
\label{eq:EccDistGauss}
\end{equation}
\end{enumerate}
In the light of these remarks, the use of the Bessel-Gaussian probability distribution 
$p_{\rm BG}(\epsilon_m)$
in (\ref{eq3.27}) does not seem to be the best motivated choice for the comparison to model
event distributions in $\epsilon_m$ (and to measured event distributions in $v_m$). The problem
with $p_{\rm BG}[w]$ is two-fold. First, the derivation of $p_{\rm BG}(\epsilon_m)$ from a 
Gaussian distribution at fixed $\phi_R$ relies on the ad hoc assumption 
$C_{l_1,l_2}^{(m_1,m_2)} = C_{l_1,l_2}^{(m_1)}\, \delta_{m_1,-m_2}$ that implies that the correlation
of fluctuations is independent of their orientation with respect to the reaction plane (see discussion of
equation (\ref{eq3.18})). Moreover, for odd $m=1,3,5,\ldots$, the Gaussian model implies 
$\bar \epsilon_m = 0$  (see eq.~(\ref{eq3.25})) and this calls into question the very form of (\ref{eq3.27}). 

In fact, for odd $m=1,3,5,\ldots$, the Gaussian model at fixed $\phi_R$ leads without further assumption
to an explicit analytical expression for $p(\epsilon_m)$ that is of Bessel-Gaussian form but that has 
arguments different from $p_{\rm BG}$. We emphasize in particular that the argument of $I_0$  
in eq.~(\ref{eq3.26}) is quadratic in $\epsilon_m$ while it is linear in  eq.~(\ref{eq3.27}). The two distributions can also be distinguished by the cumulants, see Eq.\ \eqref{eq3.34} and the discussion thereafter. The form of 
(\ref{eq3.26}) seems better motivated, as it is derived from the general form (\ref{eq3.1}) of the Gaussian 
distribution without further assumptions. For the same reason, it seems preferable to use for even 
$m=2,4,5,\ldots$ the probability distribution (\ref{eq3.20}) that depends on three parameter. 
The differences between the previously used ansatz (\ref{eq3.27}) and the expressions derived here
can be traced back to our observation (\ref{eq3.18}) that the connected two-mode correlators 
$C_{l_1,l_2}^{(m_1,m_2)}$ do not need to vanish for even and non-zero values of $m_1+m_2$.

\subsection{Distribution of flow coefficients}
\subsubsection{Linear dynamic response}
If we restrict the relation (\ref{eq2.1}) between flow coefficients $V_m$ and initial amplitudes
to the linear dynamic response, $V_m^* = S_{(m)l}\, w_{l}^{(m)}$, then we can determine
the event-by-event distribution of flow harmonics $p(v_m)$ by paralleling exactly the calculation 
of eccentricities given above,
	\begin{eqnarray}
	p(v_m) &=&  v_m \int d\phi_R 
	\int D w \; \delta^{(2)}\left(V^*_m - \sum\nolimits_{l=1}^\infty S_{(m)l} 
	w^{(m)}_l \right) \; p[w] 
	\nonumber \\
	&=& \frac{v_m}{\pi\sqrt{\kappa_m^2-\kappa_m^{\prime 2}}} \int_0^{2\pi} d \phi 
	\nonumber \\
	&& \times  \exp\left(
	-\frac{v_m^2 [ \kappa_m-\kappa_m^\prime \cos(2\phi) ] +\bar v_m^2 [\kappa_m-\kappa_m^\prime] - 2 v_m \bar v_m \cos(\phi) [\kappa_m-\kappa_m^\prime]}{\kappa_m^2-\kappa_m^{\prime 2}}
	 \right)\, ,\nonumber \\
	\label{eq3.28}
	\end{eqnarray}
where
	\begin{eqnarray}
	\bar v_m &=& 
	\sum_{l=1}^\infty S_{(m)l} \, \bar w^{(m)}_l
	\label{eq3.29}\\
	\kappa_m &=& \sum_{l_1,l_2=1}^\infty  
	S_{(m) l_1} S_{(-m)l_2} \; C^{(m,-m)}_{l_1,l_2},
	\label{eq3.30}\\
	\kappa_m^\prime &=& \sum_{l_1,l_2=1}^\infty  
	S_{(m)l_1} S_{(m)l_2} \; C^{(m,m)}_{l_1,l_2}.
	\label{eq3.31}
	\end{eqnarray}
We note that we have not made the assumption $v_m \sim \epsilon_m$ here. In contrast, we assume that both $v_m$ and $\epsilon_m$ are given as linear combinations of the Bessel-Fourier coefficients $w^{(m)}_l$. This is a weaker assumption since ${\cal K}^{(m)}_l$ in eq.\ \eqref{eq3.19} and $S_{(m) l}$ in eq.\ \eqref{eq3.29}, seen as vectors with index $l$, do not have to be parallel. 

\begin{figure}
\centering
\includegraphics[width=0.8\textwidth]{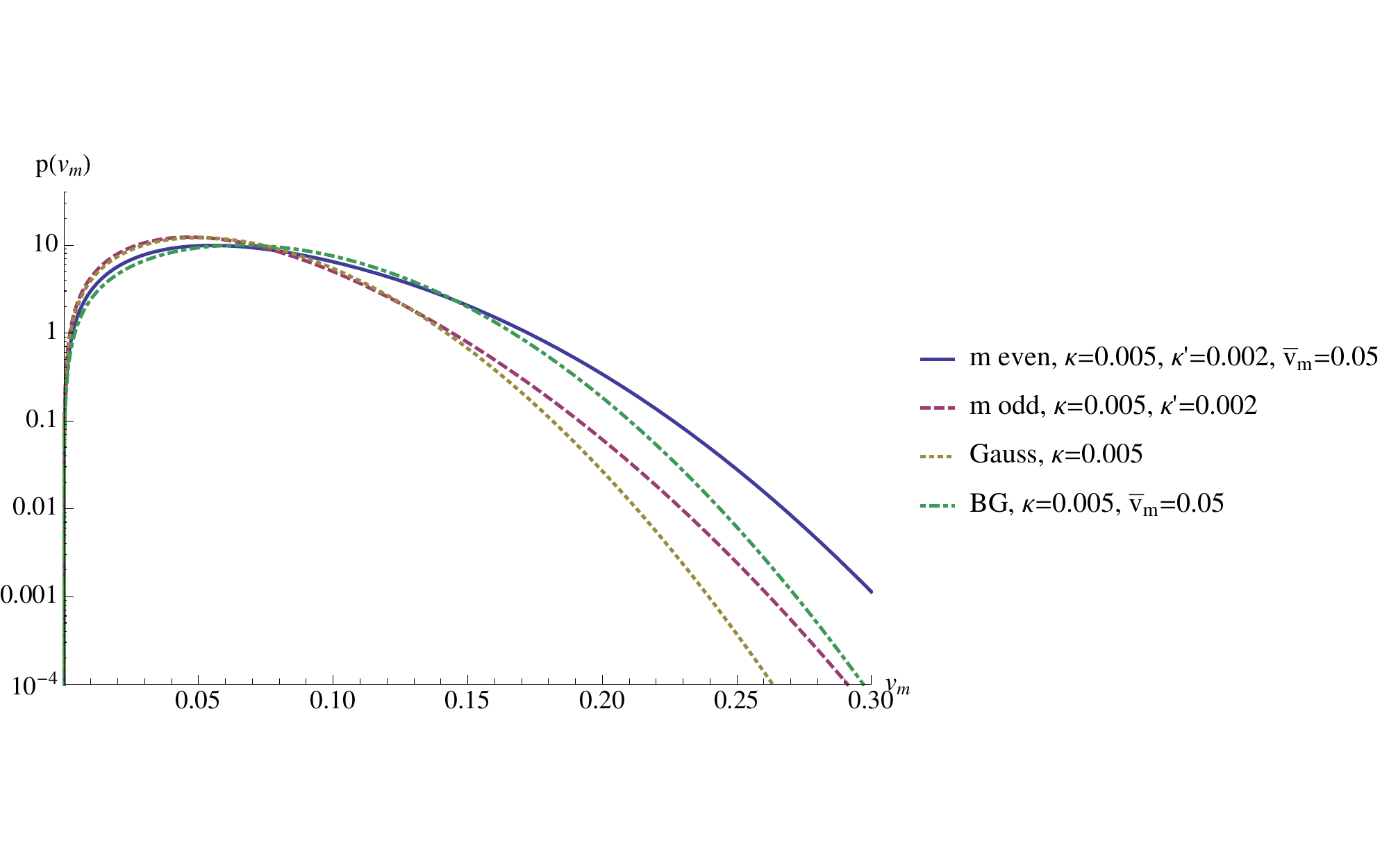}
\vspace{-1cm}
\caption{For $m$ even, the probability distribution of event-by-event flow harmonics (\ref{eq3.28}) 
depends on two variances $\kappa$, $\kappa^\prime$ and one expectation value $\bar v_m$ 
(solid line). For a choice of these parameters, the plot compares to the same distribution for $m$ odd
(dashed line, $\bar v_m$ vanishes) and to the corresponding Gaussian distribution that
results for central collisions (dotted line, $\bar v_m$ and $\kappa^\prime$ vanish).
We also compare to the Bessel-Gaussian distribution that results from the assumption $\kappa^\prime=0$
at finite $\bar v_m$.}
\label{fig1}
\end{figure}

All the remarks made above about event-by-event distributions in eccentricity carry over trivially to 
$p(v_m)$ if one restricts the discussion to linear dynamic response terms. In particular, for $m$ even,
equation (\ref{eq3.28}) depends on the three parameters $\kappa_m$, $\kappa^\prime_m$ and 
$\bar v_m$, and $p(v_m)$ has the same functional form as the eccentricity distribution (\ref{eq3.20}).
For odd $m$, reflection symmetry implies $\bar v_m = 0$, and $p(v_m)$ reduces to a two-parameter
function of the form of eq.~(\ref{eq3.26}). And at vanishing impact parameter, azimuthal symmetry
implies that $\kappa_m^\prime  = \bar v_m = 0$, and one obtains from (\ref{eq3.28}) a Gaussian, 
in complete analogy to (\ref{eq:EccDistGauss}).  Finally, the Bessel-Gaussian distribution 
proposed in ref.\ \cite{Voloshin:2007pc} is obtained by assuming $\kappa^\prime_m = 0$ but
keeping $\bar v_m$ finite. In fig.\ \ref{fig1} we compare these four distributions for one set of
parameters $\kappa_m$, $\kappa_m^\prime$ and $\bar v_m$. In this section, we have pointed
out for the first time that if one starts from initial fluctuations that follow a Gaussian distribution 
at fixed orientation of the reaction plane, then one can have a non-vanishing off-diagonal variance
$\kappa^\prime_m$ in the distribution of $p(v_m)$. Fig.\ \ref{fig1} serves to illustrate that such a small
non-vanishing  value $\kappa_m^\prime$ can affect the shape of event-by-event distributions 
in flow harmonics. 

%%%%%%%%%%%%%%%%%%%%%%%%%%%%%%%%%%%%%%%%%%%%%%%%
%%%%%%%%%%%%%%%%%%%%%%%%%%%%%%%%%%%%%%%%%%%%%%%%%%%

Event-by-event distributions $p(v_m)$ of flow harmonics were measured recently in 
Pb+Pb collisions at the LHC for  $m=2,3,4$ and for different centrality classes~\cite{Aad:2013xma}.
These measured distributions were also characterized in terms of their variance
$\sqrt{\langle v_m^2\rangle - \langle v_m\rangle^2}$, their mean $\langle v_m\rangle$,
and the ratio of these quantities  that takes the value 
\begin{equation}
	\frac{\sqrt{\langle v_m^2\rangle - \langle v_m\rangle^2}}{\langle v_m\rangle}
	= \sqrt{ \frac{4}{\pi}-1} \qquad \hbox{for Gaussian distributions.}
	\label{eq3.ratio}
\end{equation}
It was found that the distributions for $m=3$ and $4$ are within errors
consistent with (\ref{eq3.ratio}), while the distribution for $m=2$ is characterized by
a value significantly smaller than $\sqrt{ \textstyle\frac{4}{\pi}-1}$ for non-central
collisions~\cite{Aad:2013xma}. One may wonder whether the more general form of the 
probability distribution (\ref{eq3.28}) derived here can lead to an improved description 
of these data. While a comparison to data lies outside the scope of this work, we mention 
in this context that the distribution obtained from (\ref{eq3.28}) for odd $m$ satisfies
\begin{equation}
	\frac{\sqrt{\langle v_m^2\rangle - \langle v_m\rangle^2}}{\langle v_m\rangle}
	= \sqrt{ \frac{4}{\pi}-1} + \frac{1}{2\pi} \frac{{\kappa_m^\prime}^2}{\kappa_m^2} + 
	{\cal O}\left((\kappa_m^\prime)^3\right) \qquad \hbox{for $m$ odd,}
	\label{eq3.ratioc}
\end{equation}
In contrast, by setting $\kappa^\prime_m = 0$ in (\ref{eq3.28}) we find a deviation 
from the Gaussian result (\ref{eq3.ratio}) that has the opposite sign. 
\begin{equation}
	\frac{\sqrt{\langle v_m^2\rangle - \langle v_m\rangle^2}}{\langle v_m\rangle}
	= \sqrt{ \frac{4}{\pi}-1} 
	\left( 1 - \frac{\bar v_m^4}{4\, \kappa_m^2\, (4 - \pi)} \right) + 
	{\cal O}\left(\bar v_m^6\right) \qquad \hbox{for a Bessel-Gaussian distribution.}
	\label{eq3.ratiod}
\end{equation}
Therefore, depending on the choice of input parameters, the full distribution (\ref{eq3.28}) 
valid for even $m$ can show deviations from (\ref{eq3.ratio}) of either sign. 

As stated above, ATLAS data on $v_3$ are consistent with (\ref{eq3.ratio}). One therefore
requires from (\ref{eq3.ratioc}) that 
$ \textstyle \frac{1}{2\pi} \frac{{\kappa_m^\prime}^2}{\kappa_m^2} $ is small compared
to the value of $\sqrt{ \frac{4}{\pi}-1} \sim 0.52$. We caution, however, that there may be
tests of the distribution $p(v_m)$ that are more straightforward than a comparison to  
$\textstyle\frac{\sqrt{\langle v_m^2\rangle - \langle v_m\rangle^2}}{\langle v_m\rangle}$.
For instance, it follows from (\ref{eq3.12}), (\ref{eq3.13}) and the form of (\ref{eq3.28}) that
\begin{equation}
\begin{split}
v_m\{2\}^2 = & \bar v_m^2 + \kappa_m + \hbox{non-linear terms},\\
v_m\lbrace{4\rbrace}^4 = & \bar v_m^4 - 2 \kappa_m^\prime \bar v_m^2 - \kappa^{\prime 2}_m  + \hbox{non-linear terms},\\
v_m\{6\}^6 = & \bar v_m^6 - 3 \bar v_m^4 \kappa^\prime_m  + \hbox{non-linear terms},\\
v_m\{8\}^8 = & \bar v_m^8 - 4 \bar v_m^6 \kappa^\prime_m  + 2 \bar v_m^4 \kappa_m^{\prime 2} + \frac{12}{11} \bar v_m^2  \kappa_m^{\prime 3} + \frac{3}{11} \kappa_m^{\prime 4}+ \hbox{non-linear terms}.
\end{split}	
\label{eq3.34}
\end{equation}
Note that for $m$ odd, this implies in particular $v_m\lbrace{4\rbrace}^4 = -\kappa^{\prime 2}_m $ and $v_m\lbrace{6\rbrace}^6=0$. The measurements
of positive values for $v_3\lbrace{4\rbrace}^4$ and non-zero values for $v_3\lbrace{6\rbrace}^6$ thus falsify the phenomenological validity
of the distribution (\ref{eq3.28}) for $m=3$. This implies that at least one of the
two basic assumptions underlying (\ref{eq3.28}) must be wrong: the dynamical response
may not be linear and/or the distribution of initial fluctuations at fixed $\phi_R$ may not be
Gaussian. We shall comment in the next subsection on the first possibility, before exploring
in section ~\ref{sec4} in detail the case of universal deviations from a Gaussian distribution
of fluctuations. 

%%%%%%%%%%%%%%%%%%%%%%%%%%%%%%%%%%%%%%%%%%
%
\subsubsection{Non-linear dynamic response}
In principle, the role of the non-linear dynamic response terms in (\ref{eq2.1}) on the event-by-event
distributions of $v_m$ can still be discussed on the level of the probability distribution $p(v_m)$ by evaluating 
(\ref{eq3.28}) with a non-linear constraint 
\begin{equation}
\delta^{(2)}\left(V^*_m - \sum_{l_1} S_{(m)l_1} 
w^{(m)}_{l_1} - \sum_{m_1,m_2.l_1,l_2} S_{(m_1,m_2)l_1,l_2} w^{(m_1)}_{l_1} w^{(m_2)}_{l_2} \delta_{m,m_1+m_2} - \dots \right)
\end{equation}
in the argument. In practice, this evaluation has to be done numerically and is likely to be involved. 
To gain insight into the role of non-linear dynamical response terms, we therefore turn to the study
of the cumulants that characterize $p(v_m)$. Here, we make the following remarks:
\begin{enumerate}
	\item For {\it vanishing impact parameter}, the second order cumulant flow is dominated
	by the linear dynamical response 
	\begin{equation}
		v_m\lbrace{2\rbrace}^2 = S_{(m)l_1}\, S_{(-m) l_2}\, C_{l_1,l_2}^{(m,-m)} + {\cal O}(w^3)\, .
	\end{equation}
	In contrast, the connected 4-mode correlator (\ref{eq3.13}) vanishes for vanishing impact
	parameter. This implies that also the first line on the right hand side of eq.\ \eqref{eq2.7}, that gives the contribution from linear response dynamics, vanishes. In other words, $v_m\{4\}^4$ depends only on terms that 
	are proportional to some power of non-linear dynamic response terms. It is easy to see
	from (\ref{eq2.7}) that these terms are non-zero in general. For vanishing impact parameter,
	the probability distribution is Gaussian with zero mean and the correlators on the right hand side
	of (\ref{eq2.7}) that involve an odd number of modes vanish. However, there are correlators involving an even number of modes, e.\ g.\ 
\begin{equation}
\begin{split}
\langle  T_{3A} T_{3B} \rangle_\circ =  \sum_{\substack{m_2,m_3,m_4,m_5 \\ l_1,\ldots,l_6}} & S_{(m)l_1} S_{(-m_2,-m_3) l_2,l_3} S_{(m_4,m_5)l_4,l_5} S_{(-m)l_6} \, \delta_{m,m_2+m_3} \delta_{m,m_4+m_5} \\
& \times \langle  w^{(m)}_{l_1} w^{(-m_2)}_{l_2} w^{(-m_3)}_{l_3} w^{(m_4)}_{l_4} w^{(m_5)}_{l_5} w^{(-m)}_{l_6} \rangle.
\end{split}
\end{equation}
%	\begin{eqnarray}
%		\langle T_{3A}\, T_{3B} \rangle_\circ  &=& S_{[m]}^{l_1}\, S_{[m_1,m-m_1]}^{l_2,l_3}\, 
%		S_{[m_1,m-m_1]}^{\bar l_1,\bar l_2}\, S_{[m]}^{\bar l_3}
%		\nonumber \\
%		&& \times \left[ 		
%		C_{l_1,\bar l_3}^{(m,-m)} C_{l_2,\bar l_1}^{(m_1,-m_1)} C_{l_3,\bar l_2}^{(m_1-m,m-m_1)}
%		+ C_{l_1,\bar l_3}^{(m,-m)} C_{l_2,\bar l_2}^{(m_1,-m_1)} C_{l_3,\bar l_1}^{(m_1-m,m-m_1)}
%		\right]\, .
%	\end{eqnarray}
The six-point correlation function on the right hand side is non-vanishing for Gaussian distributions. There are 15 different contractions out of which only some vanish for symmetry reasons. There is no reason that the other contributions at order $w^6$ should cancel on the right hand side of eq.\ \eqref{eq2.7}. 
From this, we conclude that if one assumes a Gaussian probability distribution of fluctuations in
	the limit of vanishing impact parameter, the observation of a finite value for 
	$v_m\{4\}^4$ is an unambiguous sign of non-linear dynamic response. 
	%
%	\item For {\it linear dynamic response at finite impact parameter}, it follows from (\ref{eq3.13}) that
%	\begin{equation}
%		v_m\lbrace{4\rbrace}^4 = - \kappa^{\prime 2}_m - 2 \kappa_m^\prime \bar v_m^2 + \bar v_m^4 + \hbox{non-linear terms}.
%		\label{eq3.34}
%	\end{equation}
%	In particular, for odd $m=1,3,5,\ldots$, $v_m\lbrace{4\rbrace}^4 = -\kappa^{\prime 2}_m $
%	while a determination via the typically used Bessel-Gaussian distribution would imply
%	$v_m\lbrace{4\rbrace}^4 = \bar v_m^4 $. These comments are
%	simple corollaries of the form of the probability distribution (\ref{eq3.28}). 
	\item For the case of an {\it arbitrary, non-linear dynamical response at finite impact parameter}, we know
	that the higher-order flow cumulant $v_m\lbrace{4\rbrace}^4$ will depend on physics of two different
	origins. First, it depends on linear dynamic response to the connected
	4-point correlator 
	$\langle w^{(m)}_{l_1} w^{(-m)}_{l_2} w^{(m)}_{l_3} w^{(-m)}_{l_4} \rangle_{\circ,c}$.
	This connected 4-point correlation function is non-vanishing only due to deviations of $p_\circ$ from a Gaussian 
	probability distribution. The second source are terms that	are proportional to powers of 
	non-linear dynamic response terms. The latter are not expected to vanish in the limit of small impact parameter $b$. As one approaches more and more central collisions and if $p_\circ$ becomes Gaussian in that limit, one expects that the non-linear response terms start to dominate at some point. The linear contribution to $v_m\{n\}^n$ in Eq.\ \eqref{eq3.34} vanishes for $b\to 0$ like $b^{n \cdot m}$. In order to estimate from which centrality class the non-linear terms dominate, one would have to determine their contribution quantitatively. Finally, we remark that additional terms arise on the linear level if $p_\circ$ is not Gaussian. An example for this is provided in the following section.
\end{enumerate}

%%%%%%%%%%%%%%%%%%%%%%%%%%%%%%%%%%%%%%%%%%%%% 
\section{The independent point-sources model (IPSM)}
\label{sec4}

Essentially all realistic microscopic models of initial conditions incorporate the plausible assumption that the 
initial density distribution results from the superposition of a large number of sources that
are well-localized and therefore small compared to the system size. 
The independent 
point-sources model (IPSM) realizes this assumption in a setting in which all 
correlation functions of initial fluctuations are analytically calculable, including their radial dependence, see below. Eccentricities have been calculated in this setting to various orders in $1/N$ where $N$ is the number of sources \cite{Bhalerao:2006tp, Alver:2008zza, Bhalerao:2011bp, Yan:2013laa, Bzdak:2013raa}. As emphasized 
recently by Ollitrault and Yan~\cite{Yan:2013laa} as well as Bzdak and Skokov \cite{Bzdak:2013raa}, $n$-mode correlators calculated in the IPSM show
characteristic deviations from a Gaussian distribution even at vanishing impact parameter.
Remarkably, these deviations from a Gaussian distribution display universal properties 
that are shared by the class of currently explored phenomenologically relevant 
models of initial conditions~\cite{Yan:2013laa}. This motivates us to explore in this section the properties of the IPSM in more detail. In particular, we shall extend the discussion of the IPSM
to the case of finite impact parameter when the parametric counting of $n$-mode correlators will
be seen to be different, we shall extend the discussion from a Gaussian to an arbitrary transverse 
density distribution, and we shall extend it from the characterization of eccentricity moments to
arbitrary correlators of the modes $w_l^{(m)}$ evaluated for event samples at fixed and at randomly
oriented reaction plane. 
 
\subsection{$1/N^{n-1}$ scaling for fixed reaction plane orientation}
In the IPSM, the transverse enthalpy density $w(\vec{x})$ of a particular event
is defined as a linear superposition of contributions from $N$ point sources,~\footnote{It is relatively easy to generalize the model to situations where the number of contributing sources is itself fluctuating or where the contribution of each point source fluctuates in strength. Also extended sources can be treated. All these modifications do not change the feature that $w(\vec x)$ is a superposition of independently and identically distributed random variables with a certain distribution.}
\begin{equation}
w(\vec x) =  \left[ \frac{1}{\tau_0} \frac{dW_\text{BG}}{d\eta}\right]
 \frac{1}{N} \sum_{j=1}^N \delta^{(2)}(\vec x - \vec x_j)\, .
 % ,\qquad W \equiv  \frac{1}{\tau_0} \frac{dW_\text{BG}}{d\eta}\, .
\label{eq4.1}
\end{equation}
Here, $\textstyle \frac{1}{\tau_0} \frac{dW_\text{BG}}{d\eta}$ is the event-averaged enthalpy per unit
rapidity at initial time $\tau_0$.
% rapidity $\eta$, $W \equiv \frac{1}{\tau_0} \frac{dW_\text{BG}}{d\eta}$.
The source positions $\vec x_j$ are random two-dimensional vectors that follow the same
probability distribution $p(\vec x_j)$ for all $j$. This probability distribution is normalized,
\begin{equation}
\int d^2 x \, p(\vec x) = 1.
\label{eq4.2}
\end{equation}
At fixed impact parameter,
event-by-event fluctuations in the positions $ \vec x_j$ are the only source of fluctuations in the 
IPSM. The probability distribution $p(\vec x)$ is azimuthally asymmetric as a function of impact
parameter and it becomes azimuthally symmetric for an ensemble of central events. The transverse
profile of $p(\vec x)$ determines the expectation value of the enthalpy density for an ensemble
of collisions with fixed orientation of the reaction plane 
\begin{equation}
\langle w(\vec x) \rangle = \left[ \frac{1}{\tau_0} \frac{dW_\text{BG}}{d\eta}\right] \,  p(\vec x) .
\label{eq4.3}
\end{equation}
For the calculation of $n$-point correlation functions, it is useful to introduce the partition function 
\begin{equation}
Z[j] = \left\langle e^{\int d^2 x^\prime \, j(\vec x^\prime) w(\vec x^\prime)} \right\rangle\, .
\label{eq4.4}
\end{equation}
Due to the assumption that the $\vec x$ are independently and identically distributed, this partition function factorizes into a product of contributions from each source,
\begin{equation}
Z[j] =  \left[ \int d^2x \, p(\vec x) \, e^{\frac{1}{\tau_0} \frac{dW_\text{BG}}{d\eta} \frac{1}{N} j(\vec x)}\right]^N.
\label{eq4.5}
\end{equation}
Correlation functions can now be obtained as functional derivatives of $Z[j]$, for example
\begin{equation}
\langle w(\vec x) w(\vec y) \rangle = \frac{\delta^2}{\delta j(\vec x) \delta j(\vec y)} Z[j]{\big |}_{j=0}
= \left[ \frac{1}{\tau_0} \frac{dW_\text{BG}}{d\eta}\right]^2 \left(\tfrac{1}{N} p(\vec x) \delta^{(2)}(\vec x-\vec y) + \left(1-\tfrac{1}{N}\right) p(\vec x) p(\vec y)  \right)\, .
\label{eq4.6}
\end{equation}
Similarly, connected correlation functions can be obtained from functional derivatives of $\ln Z[j]$, for example
\begin{equation}
\begin{split}
\langle w(\vec x) w(\vec y) \rangle_c = &
\langle w(\vec x) w(\vec y) \rangle - \langle w(\vec x) \rangle \langle w(\vec y) \rangle = \frac{\delta^2}{\delta j(\vec x) \delta j(\vec y)} \ln Z[j]{\big |}_{j=0} \\
= & \left[ \frac{1}{\tau_0} \frac{dW_\text{BG}}{d\eta} \right]^2 \frac{1}{N} \left[ p(\vec x) \delta^{(2)}(\vec x - \vec y) - p(\vec x) p(\vec y) \right]\, .
\end{split}
\label{eq4.7}
\end{equation}
Observe in particular the first term in the second line of eq.\ \eqref{eq4.7}. It has the form of a contact term which is due to the point-like shape of the sources. For more realistic source shape this term decays with $|\vec x- \vec y|$ on a length scale that is characteristic of its size. The second term in the second line of \eqref{eq4.7} can be seen as a correction to the disconnected part and is closely related to the model assumption of exactly $N$ sources. The prefactor changes when this number is allowed to fluctuate.
For the further discussion, it is useful to give also the explicit form of the 3-point correlation function
\begin{equation}
\begin{split}
\langle w(\vec x) w(\vec y) w(\vec z) \rangle = &
\frac{\delta^3}{\delta j(\vec x) \delta j(\vec y) \delta j(\vec z)} Z[j]{\big |}_{j=0} \\
= & \left[ \frac{1}{\tau_0} \frac{dW_\text{BG}}{d\eta} \right]^3 {\Big [} \frac{1}{N^2}  p(\vec x) \delta^{(2)}(\vec x - \vec y) \delta^{(2)}(\vec x - \vec z) \\
& + \frac{1-\frac{1}{N}}{N} p(\vec x)  \delta^{(2)}(\vec x - \vec y) p(\vec z) \; [3 \; \text{perm.}]
+ \left(1-\frac{3}{N}+\frac{2}{N^2}\right) p(\vec x) p(\vec y) p(\vec z) {\Big ]}. \\
\end{split}
\label{eq4.8}
\end{equation}
The corresponding connected 3-point correlation function takes the explicit form
\begin{eqnarray}
\langle w(\vec x) w(\vec y) w(\vec z) \rangle_c &= &
\frac{\delta^3}{\delta j(\vec x) \delta j(\vec y) \delta j(\vec z)} \ln Z[j]{\big |}_{j=0} 
\nonumber \\
&=&  \left[ \frac{1}{\tau_0} \frac{dW_{\rm BG}}{d\eta} \right]^3 \frac{1}{N^2} {\Big [}  p(\vec x) \delta^{(2)}(\vec x - \vec y) \delta^{(2)}(\vec x - \vec z) \nonumber \\
&& - p(\vec x)  \delta^{(2)}(\vec x - \vec y) p(\vec z) 
- p(\vec y) \delta^{(2)}(\vec y - \vec z) p(\vec x)
\nonumber \\
&& - p(\vec z) \delta^{(2)}(\vec z - \vec x)  p(\vec y)
+ 2 p(\vec x) p(\vec y) p(\vec z) {\Big ]}\, .
\label{eq4.8b}
\end{eqnarray}
Also higher $n$-mode correlation functions can be given explicitly, e.g.
\begin{equation}
\begin{split}
\langle w(\vec x) w(\vec y) w(\vec z) w(\vec u) \rangle = &
\frac{\delta^4}{\delta j(\vec x) \delta j(\vec y) \delta j(\vec z) \delta j(\vec u)} Z[j]{\big |}_{j=0} \\
= & \left[ \frac{1}{\tau_0} \frac{dW_\text{BG}}{d\eta}\right]^4  {\Bigg [} \frac{1}{N^3} p(\vec x) \delta^{(2)}(\vec x - \vec y) \delta^{(2)}(\vec x - \vec z) \delta^{(2)}(\vec x - \vec u) \\
&+\frac{1-\frac{1}{N}}{N^2} p(\vec x) \delta^{(2)}(\vec x-\vec y) \delta^{(2)}(\vec x-\vec z) p(\vec u) \; [4\;\text{perm.}] \\
& + \frac{1-\frac{1}{N}}{N^2} p(\vec x) \delta^{(2)}(\vec x- \vec y) p(\vec z) \delta^{(2)}(\vec z-\vec u) \;[3\;\text{perm.}] \\
& + \frac{1-\frac{3}{N}+\frac{2}{N^2}}{N} p(\vec x) \delta^{(2)}(\vec x-\vec y) p(\vec z) p(\vec u) \; [6 \; \text{perm.}] \\
& + \left( 1-\frac{6}{N} + \frac{11}{N^2} - \frac{6}{N^3} \right) p(\vec x) p(\vec y) p(\vec z) p(\vec u) {\Bigg ]}\, ,
\end{split}
\label{eq4.9}
\end{equation}
We use in Eqs.\ \eqref{eq4.8} and \eqref{eq4.9} a notation where not all permutations of space arguments are written down. The number in square brackets denotes how many there are.
The connected 3-point correlation function can be obtained from \eqref{eq4.8} by keeping only the terms $\sim 1/N^2$ and the connected 4-point correlation function from \eqref{eq4.9} by keeping only the terms $\sim 1/N^3$. In fact, one can show that for arbitrary $n$, the $n$-th functional derivative of $\ln Z[j]$ at $j=0$ scales with $1/N^{n-1}$. A closely related scaling was observed in Ref.~\cite{Yan:2013laa}
for eccentricity cumulants $\epsilon_m\lbrace 2n\rbrace^{2n}$ in central events. 
The connected $2$-, $3$-, and $4$-point
correlation functions in equations (\ref{eq4.7}), (\ref{eq4.8b}) and (\ref{eq4.9}) illustrate this
scaling. In the IPSM, the $1/N^{n-1}$ scaling holds
at arbitrary impact impact parameter {\it if} the event average $\langle \ldots \rangle$ is
defined with respect to a fixed orientation of the reaction plane. However, we anticipate here
that for an ensemble $\langle \ldots \rangle_\circ$ with randomized reaction plane, the $N$-dependence
of connected $n$-mode correlation functions does not follow this scaling (see discussion of eq.~(\ref{eq4.18})). 

%%%%%%%%%%%%%%%%%%%%%%%%%%%%%%%%%%%%%%%%%%%%%%%%%%
\subsection{Bessel-Fourier coefficients in the IPSM}
\label{sec4.2}
So far, we have derived $n$-point correlation functions of $w(\vec x)$ in position space. 
Similar to our discussion of the Gaussian model in section~\ref{sec3}, it is useful to consider 
the Bessel-Fourier transformation (\ref{eq3.4}). The entire information about $n$-point 
correlation functions, that are functions of $n$ continuous variables $\vec x_j$, is then
encoded in the $n$-mode correlators $\langle w_{l_1}^{(m_1)} \ldots  w_{l_n}^{(m_n)} \rangle$
that are sets of complex-valued numbers. As shown in section~\ref{sec2}, these $n$-mode correlators
specify the information about initial conditions that enters flow measurements. 

To write the Bessel-Fourier transform, we start from equation (\ref{eq4.3}) for the average 
enthalpy density in an event ensemble with fixed orientation of the reaction plane. From this,
one finds the corresponding average enthalpy density for 
an event sample with randomized orientation of the reaction plane,
\begin{equation}
w_\text{BG}(r) = \frac{1}{\tau_0} \frac{d W_\text{BG}}{d\eta} \langle  p(r,\phi) \rangle_\circ
= \frac{1}{2\pi} \int_0^{2\pi} d \phi_R \, \bar w(r,\phi).
\label{eq4.10}
\end{equation}
The event-averaged enthalpy density $\bar w(r,\phi)$ can then be written as a Bessel-Fourier
expansion of the form of eq.~(\ref{eq3.2}), where the coefficients $\bar w^{(m)}_l$ are determined
from the orthogonality relation, 
\begin{equation}
\bar w^{(m)}_l = \frac{\tau_0}{\frac{dW_\text{BG}}{d\eta}\left[ J_{m+1}\left(z^{(m)}_l\right) \right]^2} \int_0^{2\pi} d \phi \int_0^\infty d r \, r \left[ \bar w(r, \phi) - w_\text{BG}(r) \right] e^{-im\phi} J_m\left(z^{(m)}_l \rho(r)\right).
\label{eq4.11}
\end{equation}
For an event average with fixed orientation of the reaction plane, the expectation values
read now 
\begin{equation}
\langle w^{(m)}_l \rangle = \bar w^{(m)}_l e^{-i m \phi_R}.
\label{eq4.12}
\end{equation}
The dimensionless and real coefficients $\bar w^{(m)}_l$ depend on centrality and 
they vanish in the limit of ultra-central collisions when the difference
$\bar w(r, \phi) - 
w_\text{BG}(r)$ vanishes. At finite impact parameter, the discrete symmetry 
$\bar w(r,\phi) = \bar w(r,\phi+\pi)$ implies that 
\begin{equation}
	\bar w^{(m)}_l = 0\qquad \hbox{for odd $m=1,3,5,\ldots$}\, .
	\label{eq4.13}
\end{equation}
With the help of the orthogonality relation (\ref{eq4.11}), we can obtain from
equation (\ref{eq4.7}) the connected 2-mode correlator
\begin{equation}
\begin{split}
\langle w^{(m)}_l w^{(m^\prime)}_{l^\prime} \rangle_c =  & \frac{1}{N} {\Bigg [} \frac{1}{2}  \delta_{m+m^\prime,0}  \; b^{(m,m^\prime)}_{l,l^\prime} \\
& \quad\quad+ \sum_{\hat l = 1}^\infty \bar w^{(m+m^\prime)}_{\hat l} \frac{[ J_{m+m^\prime+1}(z^{(m+m^\prime)}_{\hat l}) ]^2}{4}   b^{(m,m^\prime, -m-m^\prime)}_{l,l^\prime,\hat l} e^{-i (m+m^\prime) \phi_R} {\Bigg ]} \\
& - \frac{1}{N} \left[
\delta_{m,0} \, b^{(0)}_l + \bar w^{(m)}_l e^{-i m \phi_R}
 \right] \left[
\delta_{m^\prime,0} \, b^{(0)}_{l^\prime} + \bar w^{(m^\prime)}_{ l^\prime} e^{-i m^\prime \phi_R} \right].
\end{split}
\label{eq4.14}
\end{equation}
The symbols $b^{(m_1,\ldots,m_n)}_{l1,\ldots,l_n}$ are defined in appendix \ref{appc}.
They are numbers defined in terms of integrals over products of Bessel functions.
For non-central collisions, the connected part of the two-point function in eq. \eqref{eq4.15} gets supplemented by a disconnected part,
\begin{equation}
\begin{split}
\langle w^{(m)}_l w^{(m^\prime)}_{l^\prime} \rangle = & \langle w^{(m)}_l w^{(m^\prime)}_{l^\prime} \rangle_c +\langle w^{(m)}_l \rangle \langle w^{(m^\prime)}_{l^\prime} \rangle \\
= & \langle w^{(m)}_l w^{(m^\prime)}_{l^\prime} \rangle_c  + \bar w^{(m)}_l \bar w^{(m^\prime)}_l e^{-i(m+m^\prime)\phi_R} .
\end{split}
\label{eq4.15}
\end{equation}
We pass now from averages $\langle \ldots \rangle$ for event ensembles with fixed orientation 
of the reaction plane to averages $\langle \ldots \rangle_\circ$ for ensembles with randomized
orientation of $\phi_R$, $\langle \ldots \rangle_\circ \equiv {\textstyle \frac{1}{2\pi} } \int d\phi_R\, 
\langle \ldots \rangle$. We find from (\ref{eq4.15}) 
\begin{eqnarray}
\langle w^{(m)}_l w^{(m^\prime)}_{l^\prime} \rangle_\circ = \langle w^{(m)}_l w^{(m^\prime)}_{l^\prime} \rangle_{\circ,c} &=&
 \langle w^{(m)}_l w^{(m^\prime)}_{l^\prime} \rangle_{c,\circ} + 
 \bar w^{(m)}_{l}\,  \bar w^{(m^\prime)}_{l^\prime} \delta_{m+m^\prime,0}
 \nonumber \\
&=& \frac{1}{N} {\Bigg [} \frac{1}{2}  \delta_{m+m^\prime,0}  \; b^{(m,m^\prime)}_{l,l^\prime} {\Bigg ]}
- \frac{1}{N} {\Bigg [}\delta_{m,0} \, \delta_{m^\prime,0} \, b^{(0)}_l b^{(0)}_{l^\prime} {\Bigg ]}
\nonumber \\
&& + \left(1-\frac{1}{N}\right) \bar w^{(m)}_l \bar w^{(m^\prime)}_{l^\prime}  \delta_{m+m^\prime,0}.
\label{eq4.16}
\end{eqnarray}
Based on these calculations, we make the following comments and observations:
\begin{enumerate}
 \item {\it $1/N^{n-1}$-scaling  is broken at finite impact parameter for event samples 
 $\langle \ldots \rangle_\circ$ with randomized orientation of the reaction plane.}\\
 In general, this follows form the fact that the operation of passing from moments to connected
 $n$-mode correlators does not commute with the operation of randomizing $\phi_R$, that means
 \begin{equation}
 	\langle \ldots \rangle_{\circ,c} \not=  \langle \ldots \rangle_{c,\circ}\, .
	\label{eq4.18}
 \end{equation}
 The simplest illustration of this fact is provided by the connected two-mode correlator (\ref{eq4.16}). 
 For modes with even $m=\pm 2, \pm 4, \ldots$ when $\bar w_l^{(m)}$ does not vanish, the connected
 two-mode correlator $\langle w^{(m)}_l w^{(m^\prime)}_{l^\prime} \rangle_{\circ,c}$ for a randomized
 ensemble contains a term that is ${\cal O}(1)$ and that thus deviates from the ${\cal O}(1/N)$ scaling of 
 $\langle w^{(m)}_l w^{(m^\prime)}_{l^\prime} \rangle_{c}$. (We note as an aside that for the specific 
 2-mode correlator (\ref{eq4.16}), ${\cal O}(1/N)$ scaling holds for $m$ odd. However, as shown later,
 this is not necessarily the case for higher connected correlators of odd modes.)
 \item {\it For small impact parameter, the $b$-dependence of $1/N^{n-1}$-scaling breaking terms can be given 
explicitly.}\\
The $1/N^{n-1}$-scaling is restored for central collisions when the
event averages $\langle \ldots \rangle_\circ$ and $\langle \ldots \rangle$ become identical. 
As a consequence, the terms that break $1/N^{n-1}$-scaling must vanish in the limit of vanishing impact parameter.
Remarkably, as explained in appendix~\ref{appc}, the powerlaw dependence with which these
terms vanish can be determined analytically for small impact parameter. In particular, the relevant term
in the connected 2-mode correlator (\ref{eq4.16}) scales like 
$w^{(m)}_l \bar w^{(m^\prime)}_{l^\prime} \propto b^{\vert m\vert + \vert m^\prime\vert}$. 
Here, the dimensionless scale is set by the system size that we identify roughly with the
Woods-Saxon diameter $D_{\rm WS}$.  The term that breaks $1/N^{n-1}$-scaling is then of
order $(b/D_{\rm WS})^{\vert m\vert + \vert m^\prime\vert}$ and $1/N^{n-1}$-scaling
is effectively restored if the impact parameter is sufficiently small such that
$(b/D_{\rm WS})^{\vert m\vert + \vert m^\prime\vert}$ becomes comparable
to $1/N$. For instance, for $m=-m^\prime =2$ and  $D_{\rm WS} \sim 10$ fm, this is the case for $b = 3$ fm (assuming $N \sim {\cal O}(100)$ 
which is realistic for lead-lead collisions as we shall see below). This illustrates that the expansion in small $b$ is not only
of academic interest but applies to event samples for an experimentally accessible range
of impact parameter. By varying centrality and thus varying $b$, it is possible to move from samples
that satisfy $1/N^{n-1}$-scaling to samples in which this scaling is broken with known parametric 
dependence. 
\item {\it The non-vanishing off-diagonal variance $C^{(m,m)}_{l,l^\prime}$ identified for Gaussian distributions in section~\ref{sec3} and characterized by equations (\ref{eq3.15})-(\ref{eq3.18}) has a 
direct analogue in the IPSM.}\\
Comparing eq.(\ref{eq3.6}) to (\ref{eq4.15}),
one identifies $C^{(m_1,m_2)}_{l_1,l_2}$ (up to a factor $e^{-i(m_1+m_2)\phi_R}$) with the right hand side
of equation (\ref{eq4.14}). This expression satisfies the properties (\ref{eq3.15})-(\ref{eq3.18}). In particular, the properties of $\bar w_l^{(m)}$ imply that $C^{(m_1,m_2)}_{l_1,l_2}$ can be non-vanishing
for even $m_1+m_2$ only. For $m_1=m_2=m\neq 0$ one finds
\begin{equation}
C^{(m,m)}_{l_1,l_2} = \frac{1}{N} \left[
\sum_{\hat l = 1}^\infty \bar w^{(2m)}_{\hat l} \frac{[ J_{2m+1}(z^{(2m)}_{\hat l}) ]^2}{4}   b^{(m,m, -2m)}_{l_1,l_2,\hat l}   -  \bar w^{(m)}_{l_1} 
 \bar w^{(m)}_{ l_2}  \right].
 \label{eq4.19}
\end{equation}
The properties of
$\bar w_l^{(m)}$ imply that (\ref{eq4.19}) vanishes for vanishing impact parameter like $b^{2|m|}$. 
We conclude that a finite value of $C^{(m,m)}_{l_1,l_2}$ is not only allowed by symmetry considerations, but it is actually of the same parametric order $O(1/N)$ as  $C^{(m,-m)}_{l_1,l_2}$ in an
explicit model of the initial conditions. 
\item {\it The short hands $b^{(m_1,\ldots,m_n)}_{l1,\ldots,l_n}$ that appear in the results
for $n$-mode correlators do not depend on details of the
collisons geometry.}\\
In fact, as seen in appendix~\ref{appc}, these symbols are simply real numbers corresponding to certain integrals over Bessel functions and independent of $\bar w^{(m)}_l$ and of $w_\text{BG}(r)$. They are therefore independent of 
the impact parameter, and they are independent of the azimuthal orientation of the collision. 
In $n$-mode correlators, they appear multiplied with a characteristic dependence in the number of 
sources $N$. 
\item {\it Defining the Bessel expansion with $\rho(r)$ in terms of background density coordinates
has technical advantages.}\\
In this work, we define the Bessel-Fourier decomposition (\ref{eq3.4}) with the help of the
function $\rho(r)$ that maps $r\in [0,\infty]$ monotonously to the range $[0,1]$ and that we
define in appendix~\ref{appb}. This is different from previous works where we used
$\rho(r) = r/R$, $R$ fixed, and where we denoted the weights $\tilde{w}_l^{(m)}$ of the corresponding 
Bessel-Fourier expansion by a tilde. The new choice has various technical advantages. In particular, all basis
functions in the sum of equation (\ref{eq3.2}) approach zero smoothly for $r\to \infty$ corresponding to $\rho \to 1$. 
Also, remarkably, in this representation there is no further dependence on $w_{\rm BG}(r)$
in the result for the 2-mode correlator (\ref{eq4.14}). This statement generalizes to all
higher $n$-mode correlators in the IPSM. Thus, information about initial geometry enters
these results only via $\bar w_l^{(m)}$. Finally, for the connected part of (\ref{eq4.16}), we
have (see appendix~\ref{appc})
\begin{equation}
b^{(m,-m)}_{l_1,l_2} = \delta_{l_1 l_2} \frac{2 \, (-1)^m }{\left[ J_{m+1}(z^{(m)}_{l_1}) \right]^2}.
\label{eq4.18}
\end{equation}
This implies in particular that at vanishing impact parameter,
$\langle w_l^{(m)} w_{l^\prime}^{(m^\prime)} \rangle_\circ$ is diagonal if viewed as
a matrix in $l$, $l^\prime$. It follows directly that in the IPSM the linear dynamic contribution to the
second order flow cumulants take the simple explict form
\begin{equation}
v_m\lbrace 2\rbrace^2 = \frac{1}{N} \sum_l  \frac{S_{(m)l}^2}{\left[ J_{m+1}(z^{(m)}_{l}) \right]^2}
+ \left(1 - \frac{1}{N} \right)   \left( \sum_l S_{(m)l}\, \bar w_l^{(m)} \right)^2\, .
\end{equation}
\item {\it Higher $n$-mode correlators of Bessel-Fourier coefficients can be given explicitly.}\\
While expressions for higher $n$-mode correlators are more lengthy, the same techniques
shown here for two-mode correlators allow one to find explicit expressions for correlators
of more than two modes. To illustrate this point, we give here the 3-mode correlator 
after azimuthal averaging over angles has been performed,
\begin{equation}
\begin{split}
\langle w^{(m_1)}_{l_1} w^{(m_2)}_{l_2} w^{(m_3)}_{l_3} \rangle_\circ  = & \delta_{m_1+m_2+m_3,0}  \;  {\Bigg [}  \frac{1}{4 N^2} \, b^{(m_1,m_2,m_3)}_{l_1,l_2,l_3}  \\
& - \frac{1}{2  N^2} \delta_{m_3,0} \; b^{(m_1,m_2)}_{l_1,l_2}  \; b^{(0)}_{l_3} \; [3\; \text{perm.}]\\
&+ 2 \frac{1}{N^2} \; \delta_{m_1,0} \, \delta_{m_2,0} \, \delta_{m_3,0} \,
b^{(0)}_{l_1} \, b^{(0)}_{l_2} \, b^{(0)}_{l_3} \\
& + \frac{1-\frac{1}{N}}{4\, N^2} \sum_{\hat l = 1}^\infty 
\bar w^{(m_1)}_{\hat l} \bar w^{(m_1)}_{l_1} [J_{m_1+1}(z^{(m_1)}_{\hat l})]^2 b^{(m_1,m_2,m_3)}_{\hat l,l_2,l_3}\;[3\; \text{perm.}]\\
& - \frac{2(1-\tfrac{1}{N})}{N}  \delta_{m_1,0} \, b^{(0)}_{l_1} \bar w^{(m_2)}_{l_2} \bar w^{(m_3)}_{l_3} \; [3\;\text{perm.}]\\
& +\left(1-\frac{3}{N}+\frac{2}{N^2}\right) \bar w^{(m_1)}_{l_1} \bar w^{(m_2)}_{l_2} \bar w^{(m_3)}_{l_3}
{\Bigg ]}.
\end{split}
\label{eq:threePointBessel}
\end{equation}
Here, the terms in the first three lines account for the contributions from fluctuations in the positions of point sources. Note that the first term is of very simple structure. In order for the second and third term to contribute, at least one of the $m_i$ needs to be $0$. The remaining terms on the right hand side of eq. \eqref{eq:threePointBessel} are quadratic or cubic in $\bar w^{(m)}_l$. They contribute therefore only for collisions at non-zero impact parameter.  The origin of these terms is an interplay of fluctuations and geometry. We note as an aside that for any  linear dynamical contribution 
of the form $S_{(m)l_1} \, S_{(-m)l_2} \, \langle w^{(m)}_{l_1} w^{(-m)}_{l_2}\rangle_\circ$ in a flow
measurement, one can write down a non-linear dynamical correction by inserting 
a mode with azimuthal wave number $m^\prime = 0$, leading to 
 $S_{(m)l_1} \, S_{(-m,0)l_2 l^\prime} \, \langle w^{(m)}_{l_1} w^{(-m)}_{l_2}
 w^{(0)}_{l^\prime}\rangle_\circ$. The dynamical propagation of initial fluctuations with $m^\prime = 0$
 leads to fluctuations in the azimuthally averaged single particle spectrum, a.k.a. radial flow.  
 In principle, these terms can lead thus to correlations between event-by-event fluctuations in radial
 flow and in the harmonic flow coefficients. 
 Eq.~(\ref{eq:threePointBessel}) demonstrates that the relevant 
 correlators $\langle w^{(m)}_{l_1} w^{(-m)}_{l_2} w^{(0)}_{l^\prime}\rangle_\circ$ are 
 non-vanishing for both central and non-central collisions. 
 \end{enumerate}

%%%%%%%%%%%%%%%%%%%%%%%%%%%%%%%%%%%%%%%%%%%%%%%%%
\subsection{The IPSM shares commonalities with more realistic models of initial conditions}
\label{sec4.3}

The IPSM model has been explored so far~\cite{Yan:2013laa} in calculations of
higher cumulants of eccentricities, $\epsilon_m\lbrace 2n\rbrace^{2n}$. For event distributions
at vanishing impact parameter, it was demonstrated that ratios of 
$\epsilon_m\lbrace 2n\rbrace^{2n}$ for $n=1, 2, 3, \ldots$ are universal in the sense that
they agree with the corresponding ratios of 'more realistic' models, such as models based on
MC Glauber or MC KLN initial conditions, or initial conditions obtained from the event generator 
DIPSY.\footnote{We caution that the word 'more realistic' used here alludes to a wanted property
that is difficult to define sharply. What can be stated safely is that these models are more complex
than IPSM, that they include features from some picture of microscopic interactions that give rise to 
initial conditions, and that one may hope to refine or scrutinize these pictures by calculations based
on QCD.} In the preceding subsection~\ref{sec4.2}, we have derived explicit expressions
for $2$-mode correlators that are more differential than the information contained in eccentricities
in that they resolve the radial dependence. We can now wonder whether the IPSM shares
generic features with more realistic models also on this more differential level. In general, this
question could be addressed for the entire hierarchy of connected $n$-mode correlators of 
$w^{(m)}_{l}$'s since they are all calculable analytically. Also, it could be addressed by comparing 
to several more realistic models. Such a comprehensive model study lies outside the scope of 
the present work. Here, we limit the discussion to an exploratory study of results for the $2$-mode correlator
$\langle w^{(m)}_{l} w^{(m)*}_{k} \rangle$ evaluated in the IPSM and compared to results of the
MC Glauber model.

\begin{figure}
\centering
\includegraphics[width=0.32\textwidth]{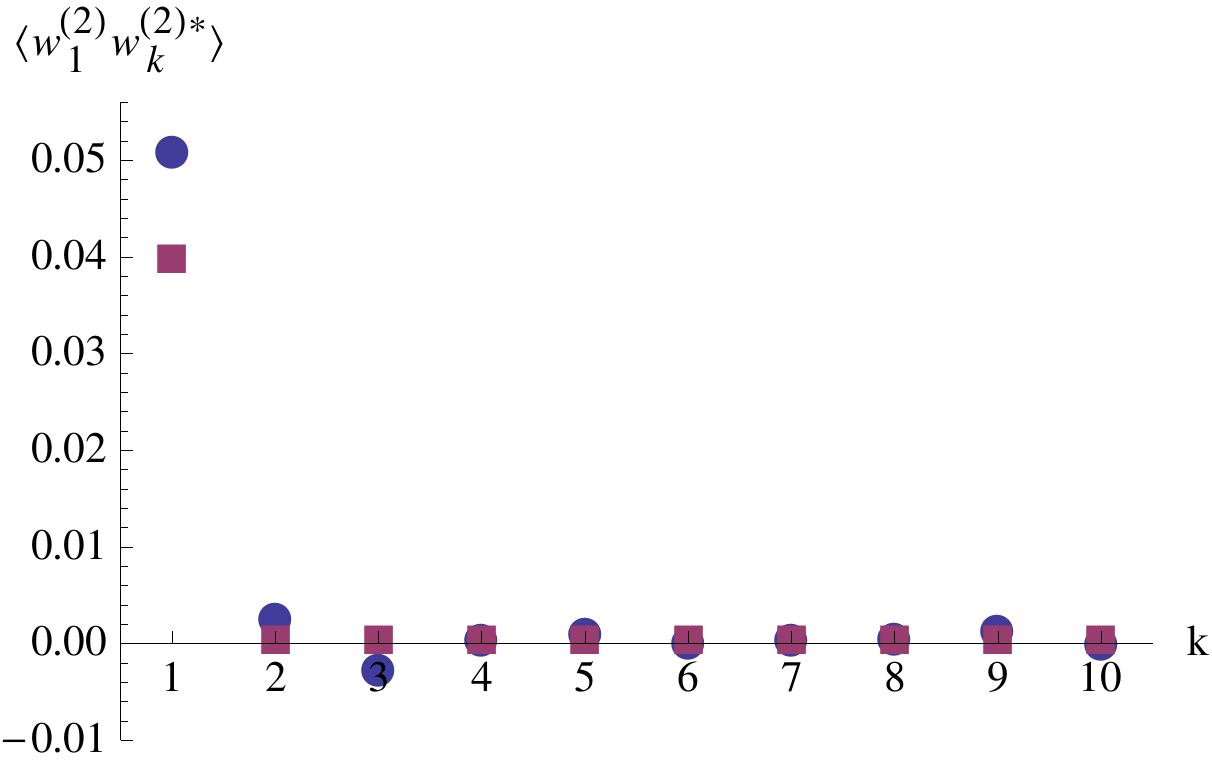}
\includegraphics[width=0.32\textwidth]{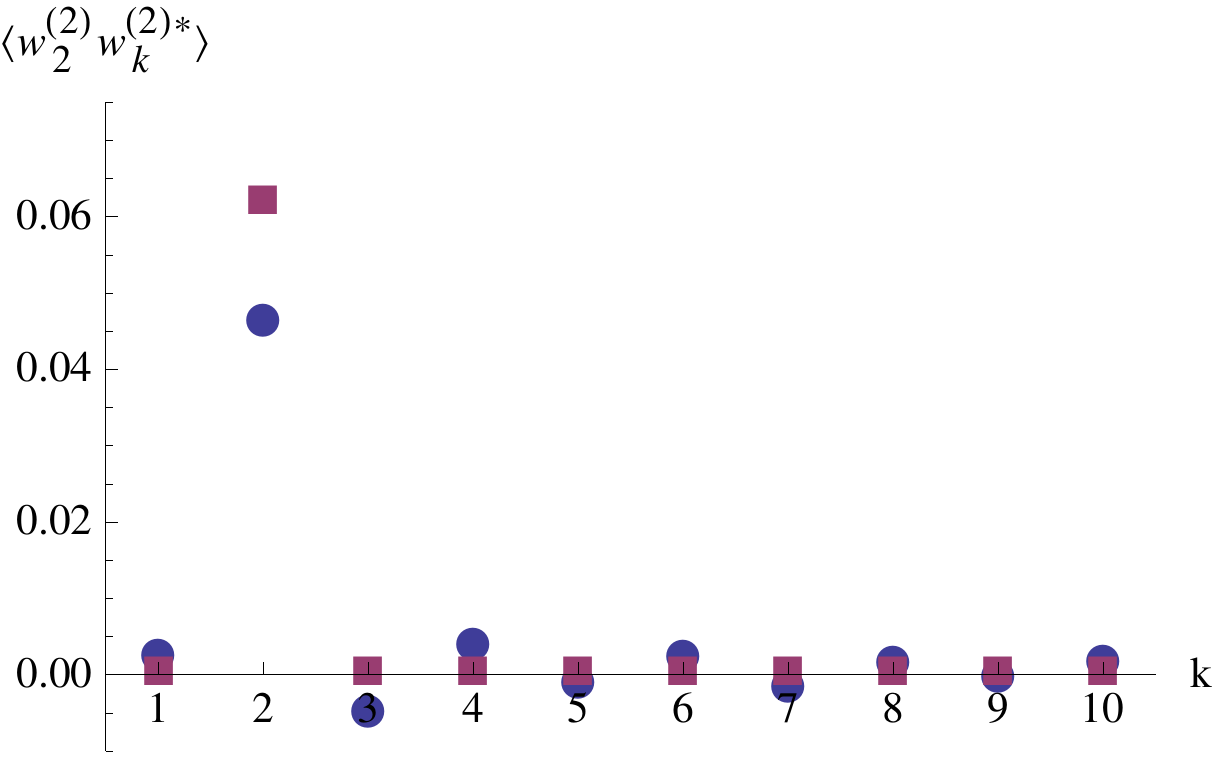}
\includegraphics[width=0.32\textwidth]{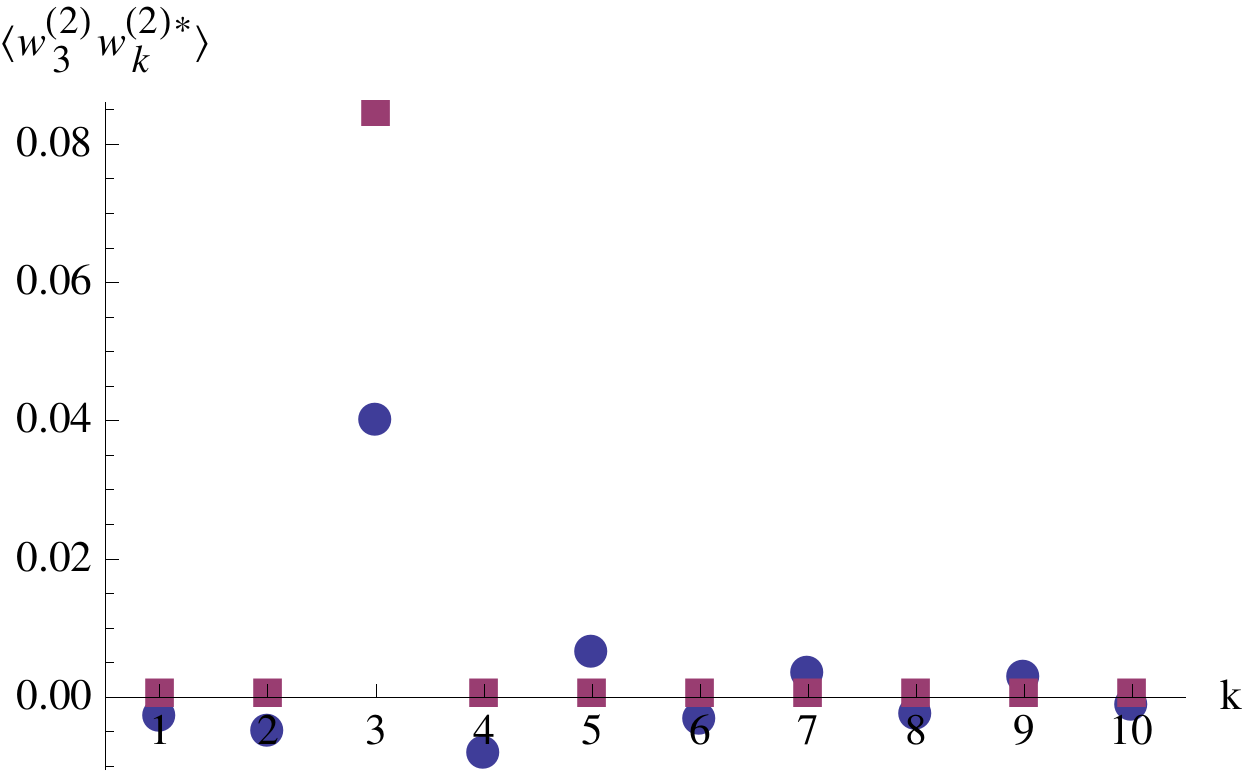}\\
\includegraphics[width=0.32\textwidth]{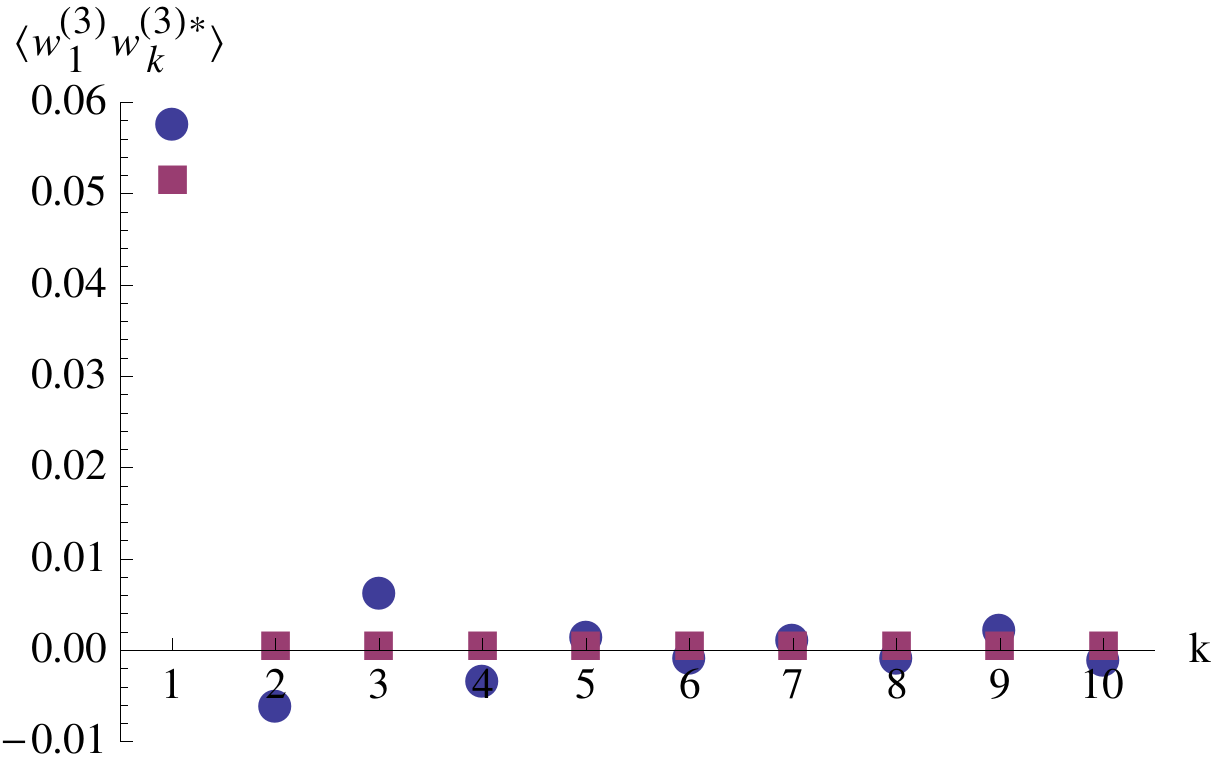}
\includegraphics[width=0.32\textwidth]{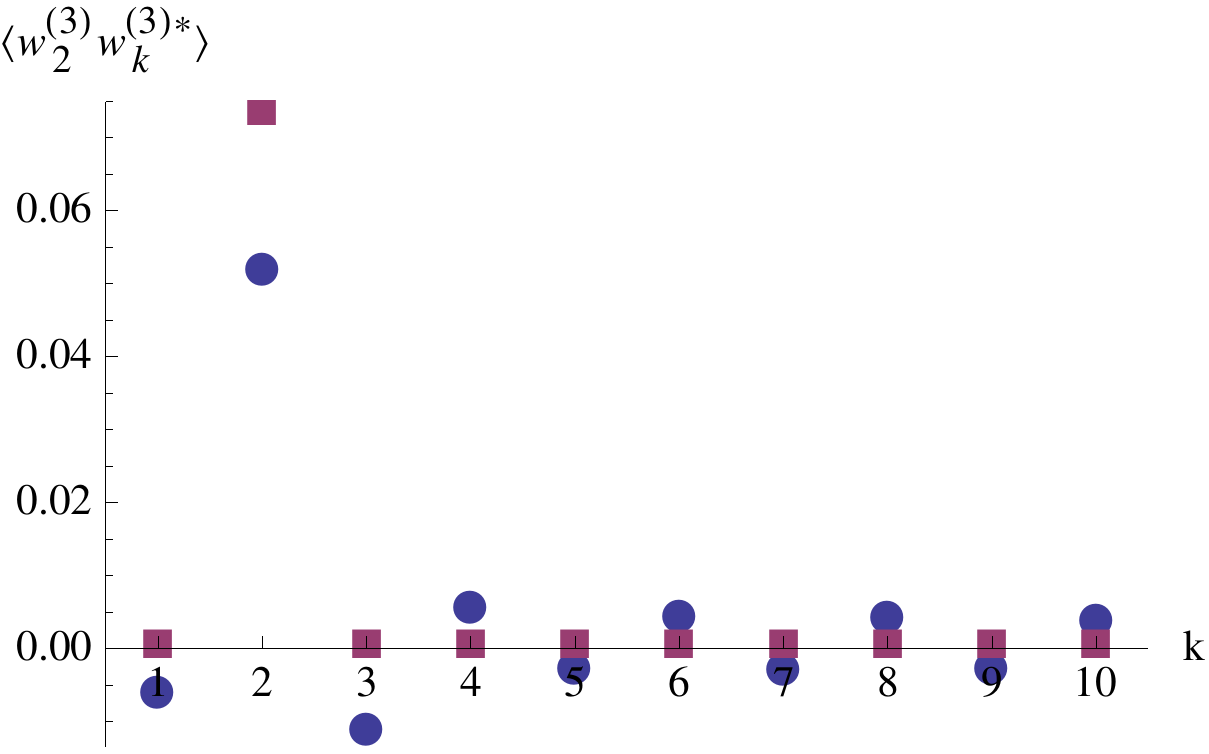}
\includegraphics[width=0.32\textwidth]{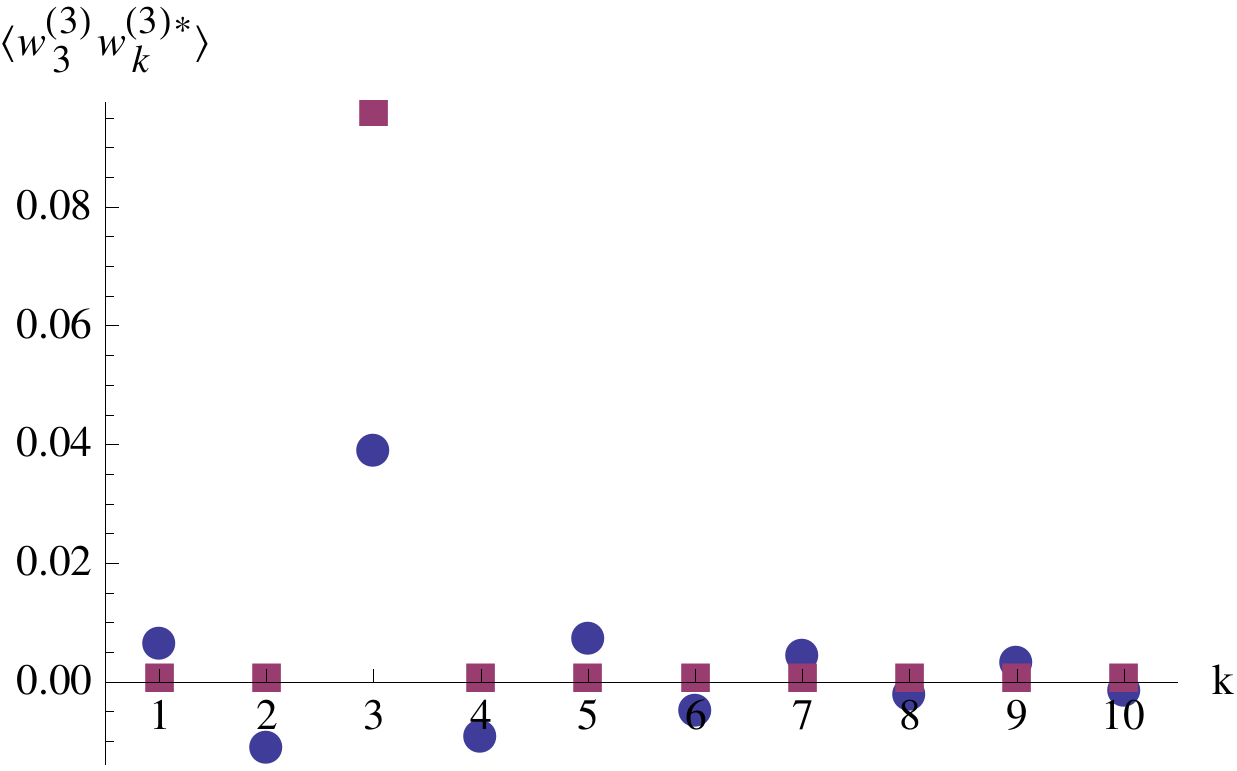}
\caption{Two-mode correlation $\langle w^{(m)}_{l} w^{(m)*}_{k} \rangle$ for $m=2$ (first line) and $m=3$ (second line) as well as $l=1$ (first column), $l=2$ (second column) and $l=3$ (third column). The points give the numerical results from a Glauber Monte-Carlo model where each participant contributes as a Gaussian source with width $\sigma=0.4$ fm. The squares give the results from the independent point source model with $N=200$ independent sources.}
\label{fig2}
\end{figure}

The specific version of the MC Glauber model that we compare to has been introduced and 
described in Ref.~\cite{Floerchinger:2013vua}. It is a MC Glauber model where each participant 
nucleon is given a Gaussian distribution in the transverse plane with width $\sigma=0.4 \,\text{fm}$.
\footnote{Ref.~\cite{Floerchinger:2013vua} gave already numerical results for the 
two-mode correlators $\langle \tilde w^{(m)}_{l} \tilde w^{(m)*}_{k} \rangle$. Here the tilde
refers to the fact that the Bessel decomposition of the enthalpy density was done
in Ref.~\cite{Floerchinger:2013vua} with a choice $\rho(r) = r/R$ in equation (\ref{eq3.2}),
while we use in the present paper the technically optimized choice for $\rho(r)$ described
in appendix~\ref{appb}. One can derive an expression for the transfer matrix
that relates the Bessel-Fourier weights of both representations, 
$w_l^{(m)} = T_{l\, l^\prime}^{(m)} \tilde w_l^{(m)}$, see appendix~\ref{appb}. 
Based on this relation, we have checked
that the data provided here for the two-mode correlator $\langle w^{(m)}_{l} w^{(m)*}_{k} \rangle$
in the MC Glauber model at vanishing impact parameter are consistent with the data on
$\langle \tilde w^{(m)}_{l} \tilde w^{(m)*}_{k} \rangle$ shown in Fig.~12 of 
Ref.~\cite{Floerchinger:2013vua}.}
Fig.\ \ref{fig2} shows the two-mode correlators $\langle w^{(m)}_{l} w^{(m)*}_{k} \rangle$ for $m=2,3$ 
and for different combinations of $l_1, l_2$. The IPSM is compared to results from the MC Glauber model for the case of collisions at vanishing impact parameter. To compare numerical results for both models, one has to fix the large parameter $N$ that appears in the IPSM. Here, we have chosen $N = 200$
on the grounds that it leads to a comparable signal strength of both models in Fig.~\ref{fig2}.
As discussed
at the end of section~\ref{sec4.2}, the correlator $\langle w^{(m)}_{l} w^{(m)*}_{k} \rangle$ is
diagonal if viewed as a matrix labeled by $l$ and $k$. Fig.~\ref{fig2} shows that off-diagonal entries
for the corresponding matrix $\langle w^{(m)}_{l} w^{(m)*}_{k} \rangle$ are non-vanishing but small
in the MC Glauber model. In this sense, the comparison of IPSM and MC Glauber model indicates
a very good albeit not perfect agreement between both models for small values of $l$. The 
deviations for larger $l$ are more pronounced. This may be explained by the fact that
larger $l$ probe finer details in position space and can therefore resolve the differences between 
a point-like and a Gaussian source shape. 

\begin{figure}
\centering
\includegraphics[width=0.4\textwidth]{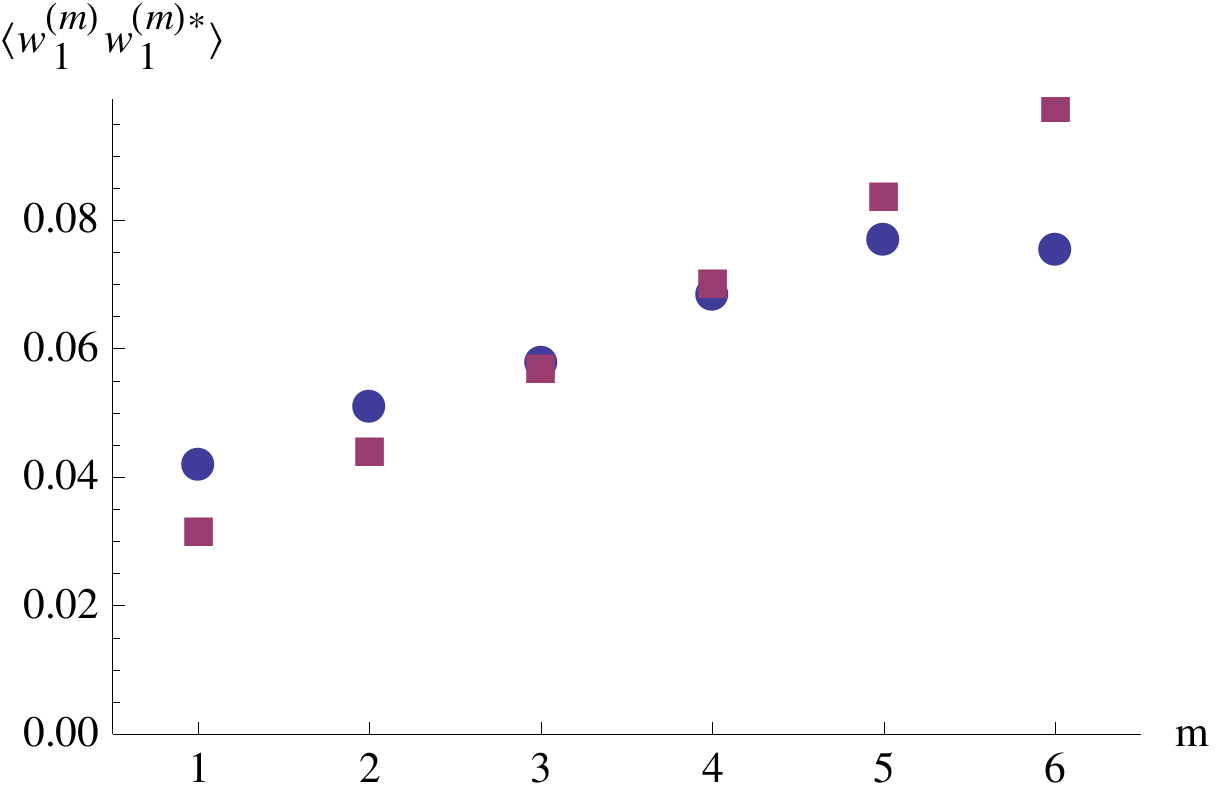}
\includegraphics[width=0.4\textwidth]{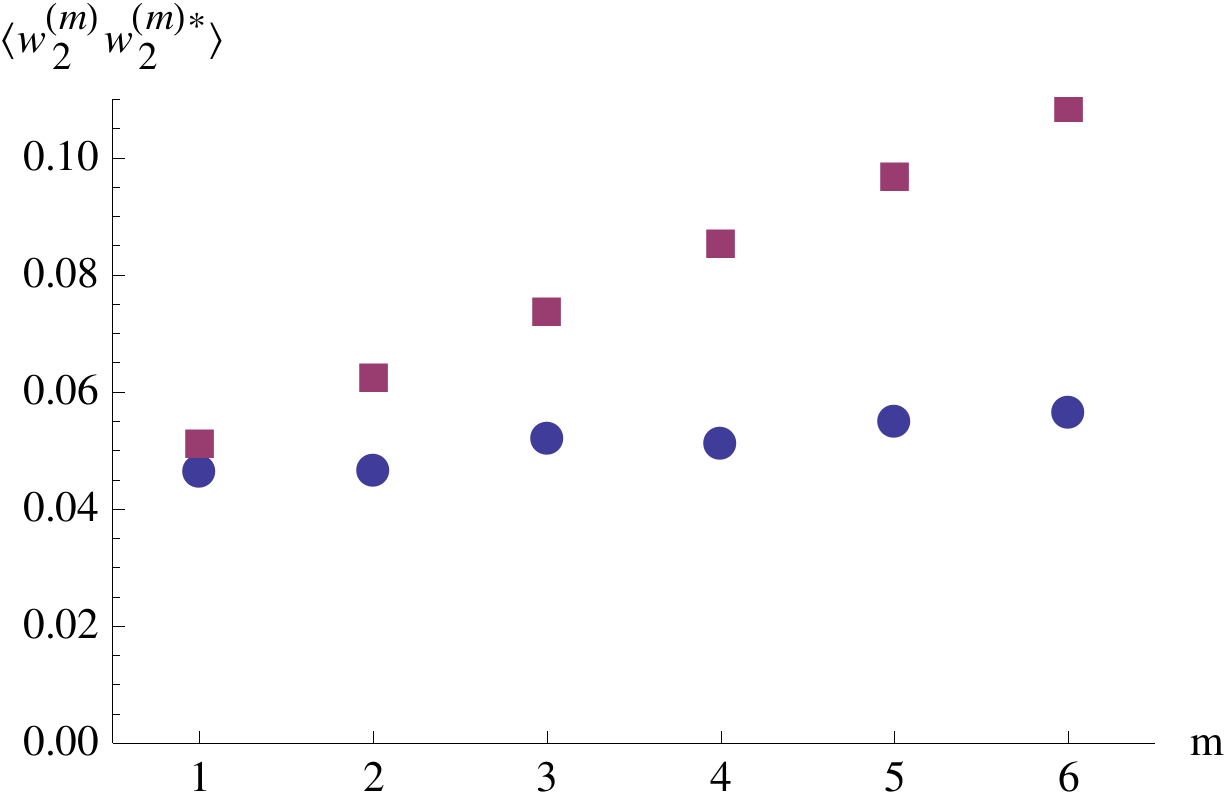}
\caption{Two-mode correlation for the radial $l=1$ mode, $\langle w^{(m)}_{1} w^{(m)*}_{1} \rangle$ (left) and the $l=2$ mode $\langle w^{(m)}_{2} w^{(m)*}_{2} \rangle$ (right) for different vales of  the azimuthal wavenumber $m$. The points give the numerical results from a Glauber Monte-Carlo model where each participant contributes as a Gaussian source with width $\sigma=0.4$ fm. The squares give the results from the independent point source model with $N=200$ independent sources.}
\label{fig3}
\end{figure}

A qualitatively similar conclusion is supported from Fig.\ \ref{fig3} that shows the diagonal terms $\langle w^{(m)}_{l} w^{(m)*}_{l} \rangle$ for radial wave number $l=1$ and $l=2$ as a function of the azimuthal wave number $m$. We compare again the Monte-Carlo Glauber model with the independent point-sources model with $N=200$. For the lowest radial mode $l=1$ corresponding to the largest wave length,
both models compare very well. But as one increases resolution in the azimuthal (i.e. increasing
$m$) or radial (i.e. increasing $l$) direction, characteristic differences between the predictions of 
the IPSM and the MC Glauber model show up. 

In summary, the qualitative agreements between both models for the results in Figs.~\ref{fig2} 
and \ref{fig3} give further support to the idea that the IPSM shares important commonalities with 
a class of more realistic models. On the qualitative level, the same results (in particular the r.h.s.
of Fig.~\ref{fig3}) give a sense of how model-specific dependencies become quantitatively
more important for higher radial and azimuthal wave numbers that resolve finer scales.  

%%%%%%%%%%%%%%%%%%%%%%%%%%%%%%%%%%%%%%%%%%%%%

\section{Concluding Remarks}
In general, a theory of experimentally measurable flow correlation measurements 
$\langle V_{m_1}\, V_{m_2} \linebreak[3]\ldots V_{m_n}\rangle_\circ$ needs to provide an understanding
for the statistics of initial density perturbations in heavy ion collisions and their fluid 
dynamical evolution. As explained in section~\ref{sec2}, this amounts to the requirement 
of knowing the initial $n$-mode 
correlators $\langle w^{(m_1)}_{l_1} \ldots w^{(m_n)}_{l_n} \rangle$ and the dynamical
response functions $S_{(m_1,\ldots,m_n) l_1,\ldots,l_n}$. 

Concerning the dynamical response functions $S_{(m_1,\ldots,m_n) l_1,\ldots,l_n}$, we know
that they depend only on the event-averaged azimuthally randomized enthalpy density  $w_{\rm BG}$
of the event class, but they do not depend on finer geometric details such as the orientation of
the reaction plane, and they do not depend on event-by-event fluctuations. An explicit method
of how to determine them without model assumptions was give in Ref.~\cite{Floerchinger:2013rya,Floerchinger:2013tya}. 
Since these dynamical response functions are (at least in principle) known, the only remaining 
model uncertainties in the calculation of flow correlation measurements are in 
determining $\langle w^{(m_1)}_{l_1} \ldots w^{(m_n)}_{l_n} \rangle$. To the extent to which
the IPSM represents universal properties shared by all realistic models of initial conditions, this
remaining model dependence is removed and model-independent predictions of fluid dynamics
become possible.\footnote{This is analogous to the situation in cosmology where
calculations of the cosmic microwave background and large scale structure can be expected to
be model independent only to the extent to which the initial conditions are constrained by
general considerations (such as symmetry arguments, based on the homogeneity and isotropy
of the system) and/or observations. The question in the present context is to what extent the IPSM
can serve a similar role in constraining initial conditions for the phenomenology of flow measurements
in heavy ion collisions.} This has motivated our detailed study of the IPSM in section~\ref{sec4}. 

In section~\ref{sec4}, we have shown how the IPSM can be solved analytically for
the full set of $n$-mode correlators $\langle w^{(m_1)}_{l_1} \ldots w^{(m_n)}_{l_n} \rangle$.
This allows to compare the IPSM quantitatively to other models on a level that is more differential
than an analysis of cumulants of eccentricities. A short comparison to a model with MC Glauber
initial conditions in section~\ref{sec4.3} has shown that the IPSM shares indeed universal
features with other models also on this more differential level, but that the analysis of 
$\langle w^{(m_1)}_{l_1} \ldots w^{(m_n)}_{l_n} \rangle$ can also serve to delineate the
azimuthal and radial length scales at which the statistics of initial perturbations in different
models shows deviations from a model-independent universal behavior. We further note that
essentially all other models of initial perturbations are defined in terms of computer codes 
that implement a physics picture. This has numerous advantages but it limits the possibilities
of finding beyond a purely numerical analysis ordering principles that explain the relative 
importance of different contributions. Comparing models to the IPSM is hence also useful
since the analytical results accessible in the IPSM allow one to find interesting ordering principles. 

In particular, we have further explored in section~\ref{sec4} the property of $1/N^{n-1}$
scaling, that is the observation that connected $n$-mode correlators 
$\langle w^{(m_1)}_{l_1} \ldots w^{(m_n)}_{l_n} \rangle_c$ in central collisions scale like
$1/N^{n-1}$ in the large parameter $N$. This is at the basis of the observation~\cite{Yan:2013laa}
that $v_m\lbrace 2n \rbrace \propto 1/N^{(2n-1)/2n}$ which explains parametrically why 
measurements of higher order flow cumulants $v_m\lbrace 2n \rbrace$ typically do not change 
within experimental errors when one increases $n$ beyond 2. We note that for central collisions,
the results of section~\ref{sec4} allow us to extend this ordering principle to a more general 
class of flow measurements. For instance, one can show that\footnote{We note that for a 
probability distribution characterized by its moments 
$M_{l_1,\dots l_n}^{(m_1,\dots m_n)}= \langle w^{(m_1)}_{l_1} w^{(m_2)}_{l_2} \ldots w^{(m_n)}_{l_n}\rangle$,
	the connected $n$-mode correlators can be written as a sum of products of moments,
\begin{equation}
	C_{l_1,l_2\dots l_n}^{(m_1,m_2,\dots m_n)}  = \sum_{\lbrace {\cal P}_n\rbrace}
	\left( \vert {\cal P}_n\ \vert - 1\right)!\, \left( -1\right)^{ \vert {\cal P}_n\ \vert - 1}
	\sum_{B \in  {\cal P}_n} \Big\langle \prod_{i\in B} w^{(m_i)}_{l_i} \Big\rangle\, .
\nonumber
\end{equation}
Here, $ {\lbrace {\cal P}_n\rbrace}$ denotes the list of all partitions of a set of size $n$,
$B \in  {\cal P}_n$ is a block in a  partition ${\cal P}_n$, and $\vert {\cal P}_n\ \vert$ 
counts the number of blocks in that partition. It follows from this structure that the
specific ${\cal O}(1/N^5)$ cumulants defined in (\ref{eq5.1}) have a different number
of non-vanishing subtraction terms on the right hand side.
	}
\begin{equation}
\begin{split}
	\langle V_{2}\, V_{3}\, V_{5}^*\,\rangle &\sim {\cal O}\left( \frac{1}{N^2}\right)\, , 
	\quad \hbox{for $b=0$,} \\
	\langle V_{2}\, V_{3}\, V_{5}^*\, V_{2}\, V_{3}\, V_{5}^*\rangle_{c}
	&= \langle \left(V_{2}\, V_{3}\, V_{5}^*\right)^2\,\rangle -  4
	\langle V_{2}\, V_{3}\, V_{5}^*\,\rangle^2 
	\sim {\cal O}\left( \frac{1}{N^5}\right) \quad \hbox{for $b=0$,}  \\
	\langle V_{2}\, V_{2}\, V_{4}^*\, V_{2}\, V_{2}\, V_{4}^*\rangle_{c}
	&= \langle \left(V_{2}\, V_{2}\, V_{4}^*\right)^2\,\rangle -  6
	\langle V_{2}\, V_{2}\, V_{4}^*\,\rangle^2 
	\sim {\cal O}\left( \frac{1}{N^5}\right) \quad \hbox{for $b=0$.} 
\end{split}
\label{eq5.1}
\end{equation}
Such measurements are interesting since they depend on $n$-mode correlators that are not
tested in the measurement of flow cumulants. We note that in these expressions, the scaling
in orders of $1/N$ applies not only to the linear response term, but also to the 
contribution of the non-linear dynamical response.\footnote{This can
be checked for the first orders of the perturbative series in equation (\ref{eq2.1}) by direct
calculation. In general, it follows from a theorem given in reference \cite{JamesMayne}.} 

At finite impact parameter, we have pointed out that flow correlation measurements with respect
to azimuthally randomized event samples cannot be ordered in powers of $1/N$. Since even the
most central event class contains events with finite albeit small impact parameter, this raises
the question to what extent the $1/N^{n-1}$ scaling in central events can be of practical use. Here,
we have shown that deviations from $1/N^{n-1}$ scaling in the IPSM show a characteristic and
analytically accessible powerlaw dependence on impact parameter. This allows one to
estimate the range of impact parameter for which terms that violate $1/N^{n-1}$ scaling are sufficiently
small to make $1/N^{n-1}$ scaling an applicable principle. Given that the impact parameter dependence
of different linear and non-linear contributions to flow correlation measurements
is different in general, one may also hope that the analytical knowledge of this $b$-dependence
can help to disentangle different dynamical contributions. However, in the present paper,
we have not yet explored this possibility further. 

We close by relating some of our results to the question of why p+Pb collisions at the LHC
show flow cumulants $v_m\lbrace 2\rbrace$, $v_m\lbrace 4\rbrace$, ($m=2,3$) that are
comparable in size and $p_T$-dependence to corresponding measurements in Pb+Pb
collisions~\cite{Aad:2013fja,Chatrchyan:2013nka}. This fact has been found in fluid dynamic 
simulations prior to the measurements~\cite{Bozek:2011if,Bozek:2012gr}, and it is currently
the focus of an important topical debate, see e.g.~\cite{Bozek:2013sda,McLerran:2013oju,Basar:2013hea}. 
One question in this context is whether a hydrodynamic explanation 
can be regarded as being generic, or whether it reproduces data only with specific 
model-dependent choices. Here, we observe that in the IPSM, the parameter $N$ can be 
viewed as increasing monotonously with the number of participants in the nuclear overlap. 
The parametric estimates for flow cumulants in pPb and PbPb read then
\begin{eqnarray}
	v_m\lbrace 2\rbrace^2\vert_\text{pPb} &\sim& \left( S_{(m)}^\text{pPb}\right)^2\, \frac{1}{N_\text{pPb}}\, ,
	\qquad v_m\lbrace 2\rbrace^2\vert_\text{PbPb} \sim \left( S_{(m)}^\text{PbPb}\right)^2\, \frac{1}{N_\text{PbPb}}\, ,
	\nonumber \\
	v_m\lbrace 4\rbrace^4\vert_\text{pPb} &\sim& \left( S_{(m)}^\text{pPb}\right)^4\, \frac{1}{N_\text{pPb}^3}\, ,
	\qquad v_m\lbrace 4\rbrace^4\vert_\text{PbPb} \sim \left( S_{(m)}^\text{PbPb}\right)^4\, \frac{1}{N_\text{PbPb}^3}\, .
	\label{eq5.3}
\end{eqnarray}
Here, we have considered only the linear dynamic response terms that we write schematically without indicating their dependence on $l$. Based on these parametric estimates, one can relate the
strength of the dynamic response to density fluctuations in different systems. For instance, 
\begin{equation}
	\left( S_{(m)}^\text{PbPb}\right) \simeq 
	\left( S_{(m)}^\text{pPb}\right) \left( \frac{N_\text{PbPb}}{N_\text{pPb}} \right)^{3/4}\, ,
	\qquad \hbox{if $v_m\lbrace 4\rbrace\vert_\text{pPb} \sim v_m\lbrace 4\rbrace\vert_\text{PbPb}$\, .}
	\label{eq5.5}
\end{equation}
There is phenomenological support for an almost linear relation between event multiplicity and the 
number of participants in a pPb or PbPb collision. Relating the number of participants approximately 
linearly to the parameter $N$ in the IPSM, one can then consider different limiting cases:
\begin{enumerate}
	\item {\it The case $N_{pPb} \simeq N_{PbPb}$ that may be realized e.g. by comparing pPb and PbPb
	collisions of similar multiplicity.}\\
	In this case, comparable flow measurements in pPb and PbPb imply comparable fluid dynamic 
	response $S_{(m)}^\text{PbPb} \simeq S_{(m)}^\text{pPb}$, see eq.~(\ref{eq5.5}). 
	\item {\it The case $N_\text{pPb} \ll N_\text{PbPb}$ that may be realized e.g. by comparing central pPb to central PbPb collisions.}\\
	In this case, for all initial conditions for which connected $n$-mode correlators of initial
fluctuations scale with $1/N^{n-1}$, the dynamic flow response $S_{(m)}$ must 
be parametrically larger for larger systems to yield harmonic flow coefficients $v_m$ that are 
independent of system size. Comparable values for 
$v_m\lbrace 4\rbrace\vert_\text{pPb}$ and $v_m\lbrace 4\rbrace\vert_\text{PbPb}$ are then
consistent with the intuitive expectation that the strength of flow phenomena increases
with system size.~\footnote{The particular parametric powerlaw dependence $\propto \left( N_\text{PbPb}/N_\text{pPb} \right)^{3/4}$ given in (\ref{eq5.5}) was obtained by requiring parametric equality of the fourth-order flow cumulants in p+Pb and Pb+Pb. If we requires parametric equality for 6-th (8-th) order flow
cumulants instead, one finds a power law $\propto \left( N_\text{PbPb}/N_\text{pPb} \right)^{\alpha}$ with $\alpha = 5/6$  ($\alpha = 7/8$).} 
\end{enumerate}
In both cases, we have obtained statements about the relative parametric strength of the dynamic
response coefficients $S_{(m)}$ in different collision systems. Note that these statements
can be tested in a fluid dynamic calculations involving only minimal model assumptions. The dynamical
response coefficients $S_{(m)}$ depend on the size of the system only via their dependence 
on the average background enthalpy $w_{\rm BG}(r)$ but they do not carry any information 
about finer details of the initial transverse density distribution. In the IPSM formulated 
in section~\ref{sec4}, $w_{\rm BG}(r)$ and the parameter $N$ can be chosen independently,
but a more complete model of the initial state and the early dynamics will relate the number
of sources $N$  to the size and to the radial dependence of the average enthalpy 
density $w_{\rm BG}(r)$. Since we know how to calculate without model-dependent assumptions
the dependence of $S_{(m)}$ on $w_{\rm BG}(r)$~\cite{Floerchinger:2013rya,Floerchinger:2013tya},
and since the relation between $w_{\rm BG}(r)$ and $N$ has only a relatively mild model dependence,
one can therefore test whether hydrodynamic evolution is consistent with 
the parametric scaling of $S_{(m)}$ required by equation~(\ref{eq5.5}). In our view, such a test could
contribute to the important question of whether fluid dynamics can account naturally for the flow
coefficients measured in systems of significantly different size, or whether some elements of
fine-tuning of initial fluctuations needs to be invoked. We plan to explore this point in the
near future. Here, we restrict us to formulating the question with the help of the results and insights
gained in section~\ref{sec4}. This is one illustration how the knowledge about the statistics of
initial density perturbations may contribute to the further understanding of
flow phenomena in nucleus-nucleus and proton-nucleus collisions.

%%%%%%%%%%%%%%%%%%%%%%%%%%%%%%%%%%%%%%%%%%%
\begin{appendix}
\section{Background density coordinates}
\label{appb}

\begin{figure}
\centering
\includegraphics[width=0.4\textwidth]{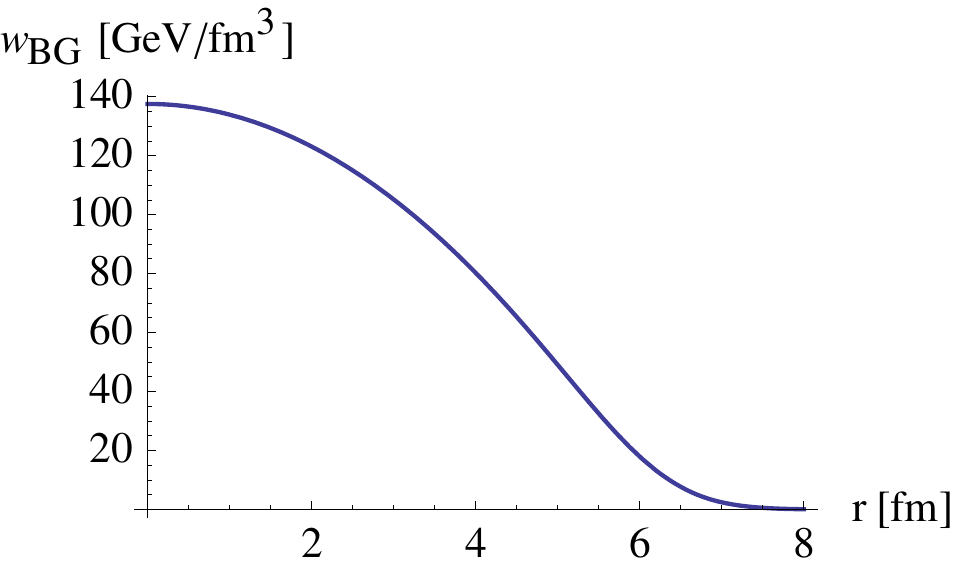}
\includegraphics[width=0.4\textwidth]{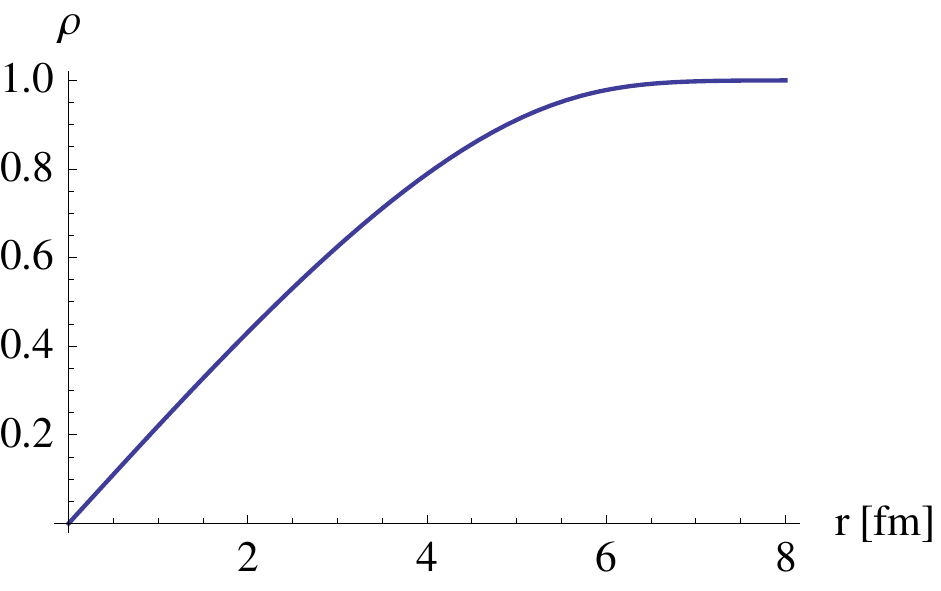}
\caption{Example for a background enthalpy distribution $w_\text{BG}(r)$ as a function of radius and the corresponding background density coordinate $\rho(r)$ as defined by eq.\ \eqref{eq:defrho}.}
\label{fig:BGCoordinate}
\end{figure}

In this appendix we discuss a special coordinate system which can be defined for a given background enthalpy density. This coordinate system is particularly well suited for the characterization of initial fluctuations and for the numerical solution of the fluid dynamic evolution equations in the background-field formalism.
We start from a transverse density distribution that is azimuthal rotation and Bjorken boost invariant,
\begin{equation}
\frac{1}{\tau_0} \frac{dW_\text{BG}}{dx_1dx_2d\eta}= \frac{1}{\tau_0} \frac{d W_\text{BG}}{r dr d\phi d\eta} = w_\text{BG}(r).
\end{equation}
Usually $w_\text{BG}(r)$ decays rather quickly with increasing $r$ outside of some radius which is of the order of a few fm. Also, the integrated enthalpy density per unit rapidity is finite,
\begin{equation}
\frac{1}{2\pi \tau_0}\frac{dW_\text{BG}}{d\eta} = \int_0^\infty dr \, r \, w_\text{BG}(r).
\end{equation}
However, there is no sharp boundary $r=R$ where the density $w_\text{BG}(r)$ goes to zero. On can define a (dimensionless) transformed coordinate $\rho(r)$ by the following relation
\begin{equation}
\rho(r) = \sqrt{\frac{\int_0^r dr^\prime \, r^\prime \, w_\text{BG}(r^\prime)}{\frac{1}{2\pi \tau_0}\frac{dW_\text{BG}}{d\eta}}} = \sqrt{\frac{\int_0^r dr^\prime \, r^\prime \, w_\text{BG}(r^\prime)}{\int_0^\infty dr^\prime \, r^\prime \, w_\text{BG}(r^\prime)}}.
\label{eq:defrho}
\end{equation}
This maps the interval $r\in (0,\infty)$ to the compact interval $\rho\in(0,1)$. For small $r$ the relation is actually linear, $\rho\sim r$ and for all $r$ the function $\rho(r)$ is monotonous. An example for a background enthalpy distribution $w_\text{BG}(r)$ and the corresponding mapping $\rho(r)$ is shown in Fig.\ \ref{fig:BGCoordinate}.

It is also useful to note the transformation behavior
\begin{equation}
\frac{\rho \, d \rho}{r \, dr} = \frac{\pi\,  \tau_0}{\frac{dW_\text{BG}}{d\eta}} w_\text{BG}(r).
\end{equation}
This implies in particular
\begin{equation}
\frac{dW_\text{BG}}{\rho d \rho d \phi d \eta} =  \frac{d W_\text{BG}}{r dr d\phi d\eta} \frac{r dr}{\rho d\rho} = \frac{1}{\pi \tau_0} \frac{d W_\text{BG}}{d \eta},
\end{equation}
which is independent of $\rho$. In other words, in the coordinate system $(\rho, \phi)$, the background enthalpy distribution is constant on the disk $\phi\in(0,2\pi)$, $\rho\in(0,1)$.

So far we have considered only the background part of the enthalpy distribution. Let us now consider an arbitrary event with fluctuations, i.e. deviations from the smooth and symmetric background part. The symmetries of the problem suggest the following expansion (we neglect a possible rapidity-dependence for simplicity)
\begin{equation}
\frac{1}{\tau_0} \frac{d W}{\rho d \rho d \phi d\eta}(\rho, \phi) = \frac{1}{\pi \tau_0} \frac{d W_\text{BG}}{d \eta}\left[ 1 + \sum_{m=-\infty}^\infty \sum_{l=1}^\infty  w^{(m)}_l e^{im\phi} J_m\left(z^{(m)}_l \rho\right) \right].
\label{eq:BFexprho}
\end{equation}
Here,  $z^{(m)}_l$ is the $l$'th zero crossing of the Bessel functions of the first kind $J_m(z)$. The coefficients $w^{(m)}_l$ can be obtained from the inverse relation
\begin{equation}
w^{(m)}_l = \frac{1}{\pi \left[ J_{m+1}(z^{(m)}_l) \right]^2} \int_0^{2\pi} d \phi \int_0^1 d \rho \, \rho\, \left[\frac{\pi}{\frac{dW_\text{BG}}{d\eta}}\frac{d W}{\rho d \rho d \phi d\eta}(\rho, \phi) - 1 \right] e^{-i m \phi} J_m\left( z^{(m)}_l \rho \right).
\label{eq:BFinvtransfrho}
\end{equation}
Within Lemoine's discrete Bessel transform approximation this reads
\begin{equation}
\begin{split}
w^{(m)}_l = & \frac{1}{2\pi} \int_0^{2\pi} d \phi \, e^{-i m \phi} \sum_{\alpha=1}^{N_\alpha} \frac{4}{\left[ z_{N_\alpha}^{(m)} J_{m+1}\left(z^{(m)}_l \right) J_{m+1}\left(z^{(m)}_\alpha \right) \right]^2} \\
& \times \left[\frac{\pi}{\frac{dW_\text{BG}}{d\eta}}\frac{d W}{\rho d \rho d \phi d\eta}\left(z^{(m)}_\alpha / z^{(m)}_{N_\alpha}, \phi\right) - 1 \right]  J_m\left( z^{(m)}_l z^{(m)}_\alpha / z^{(m)}_{N_\alpha}\right).
\end{split}
\label{eq:BFinvtransfLemoinerho}
\end{equation}
In praxis, one would replace here also the Fourier transformation by a discrete version.

When transformed back to the coordinate system $(r,\phi)$, eq.\ \eqref{eq:BFexprho} reads with $w(r,\phi) = \frac{1}{\tau_0}\frac{d W}{r d r d \phi d\eta}$
\begin{equation}
w(r, \phi) = w_\text{BG}(r) \left[ 1 + \sum_{m=-\infty}^\infty \sum_{l=1}^\infty  w^{(m)}_l e^{im\phi} J_m\left(z^{(m)}_l \rho(r)\right) \right].
\end{equation}
Note that this is equivalent to the expansion proposed in refs. \cite{Floerchinger:2013vua, Floerchinger:2013rya} except that a simpler prescription for $\rho(r)$ has been used, namely $\rho(r)=r/R$ with $R=8\,  \text{fm}$ a somewhat arbitrary radius.
Similarly, the inverse relation in eq.\ \eqref{eq:BFinvtransfrho} becomes in these coordinates
\begin{equation}
w^{(m)}_l = \frac{\tau_0}{\frac{dW_\text{BG}}{d\eta}\left[ J_{m+1}\left(z^{(m)}_l\right) \right]^2} \int_0^{2\pi} d \phi \int_0^\infty d r \, r \left[ w(r, \phi) - w_\text{BG}(r) \right] e^{-im\phi} J_m\left(z^{(m)}_l \rho(r)\right).
\label{eq:BFinvtransfr}
\end{equation}
Note that the integral on the right hand side has good convergence properties since the enthalpy density $w(r,\phi)$ decays quickly with $r$.
The discrete version according to Lemoine's method reads now ($N_\alpha$ is the number of discretization points that should be chosen larger than the maximal value of $l$ considered.)
\begin{equation}
\begin{split}
w^{(m)}_l = & \frac{1}{2\pi} \int_0^{2\pi} d \phi \, e^{-i m \phi} \sum_{\alpha=1}^{N_\alpha} \frac{4}{\left[ z_{N_\alpha}^{(m)} J_{m+1}\left(z^{(m)}_l \right) J_{m+1}\left(z^{(m)}_\alpha \right) \right]^2} \\
& \times \left[ \frac{w(r^{(m)}_\alpha,\phi)-w_\text{BG}(r^{(m)}_\alpha)}{w_\text{BG}(r^{(m)}_\alpha)} \right]  J_m\left( z^{(m)}_l z^{(m)}_\alpha / z^{(m)}_{N_\alpha}\right),
\end{split}
\label{eq:BFinvtransfLemoiner}
\end{equation}
where the radii $r^{(m)}_l$ are to be determined from the implicit relation
\begin{equation}
\rho\left(r^{(m)}_\alpha\right) = \frac{z^{(m)}_\alpha}{z^{(m)}_{N_\alpha}}.
\end{equation}
Note eq.\ \eqref{eq:BFinvtransfLemoiner} equals the expression used in refs.\ \cite{Floerchinger:2013vua, Floerchinger:2013rya} except that a simpler prescription 
\begin{equation}
r^{(m)}_\alpha = \frac{z^{(m)}_\alpha}{z^{(m)}_{N_\alpha}} R
\end{equation}
has been used there. One can also define a transfer matrix between the old and the new definition,
\begin{equation}
w^{(m)}_l = T^{(m)}_{l l^\prime} \tilde w^{(m)}_{l^\prime}
\end{equation}
with
\begin{equation}
T^{(m)}_{l l^\prime} = \sum_{\alpha=1}^{N_\alpha} \frac{4 J_m\left( \frac{z^{(m)}_l z^{(m)}_\alpha}{z_{N_\alpha}} \right) J_m\left( \frac{z^{(m)}_{l^\prime}}{R} r^{(m)}_\alpha \right)}{\left[z^{(m)}_{N_\alpha} J_{m+1}(z^{(m)}_l) J_m(z^{(m)}_\alpha)\right]^2}\, .
\end{equation}

\section{Bessel functions and integrals}
\label{appc}

In this appendix we compile some properties of Bessel functions and integrals involving them. We are particularly interested in finite integrals on the domain $\rho\in (0,1)$.

For a given set of azimuthal wave numbers $(m_1,m_2, \ldots, m_n)$ with $m_1+m_2+\ldots+m_n=0$, we introduce the following symbol
\begin{equation}
b^{(m_1,\ldots,m_n)}_{l_1,\ldots, l_n} = \frac{2^n}{\left[ J_{m_1+1}(z^{(m_1)}_{l_1}) \cdots J_{m_n+1}(z^{(m_n)}_{l_n})\right]^2} \int_0^1 d\rho \, \rho\, \left\{ J_{m_1}(z^{(m_1)}_{l_1}\rho) \cdots J_{m_n}(z^{(m_n)}_{l_n}\rho) \right\}.
\label{eqc.1}
\end{equation}
It is clear from the definition that the $b^{(m_1,\ldots,m_n)}_{l_1,\ldots, l_n}$ are symmetric with respect to the interchange of any pair of indices, e.g. $b^{(m_1,m_2)}_{l_1,l_2} = b^{(m_2,m_1)}_{l_2,l_1}$. 

We now discuss the simplest cases of $n=1,2$ where one can obtain analytic expressions. For $n=1$ there is only the possibility of $m=0$,
\begin{equation}
b^{(0)}_l = \frac{2}{\left[ J_1(z^{(0)}_{l}) \right]^2} \int_0^1 d\rho \, \rho\, J_{0}(z^{(0)}_{l}\rho) = \frac{2}{ z^{(0)}_l J_1(z^{(0)}_{l}) }.
\end{equation}
For $n=2$ one has $m_1=-m_2=m$ and obtains, using $J_{-m}(z) = (-1)^m J_m(z)$ and the orthogonality property of the Bessel functions,
\begin{equation}
\begin{split}
b^{(m,-m)}_{l_1,l_2} = & \frac{4 \, (-1)^m}{\left[ J_{m+1}(z^{(m)}_{l_1}) J_{m+1}(z^{(m)}_{l_2}) \right]^2} \int_0^1 d\rho \, \rho\, \left\{ J_{m}(z^{(m)}_{l_1}\rho) J_{m}(z^{(m)}_{l_2}\rho) \right\}\\
= &\delta_{l_1 l_2} \frac{2 \, (-1)^m }{\left[ J_{m+1}(z^{(m)}_{l_1}) \right]^2}.
\end{split}
\label{eq:bmmll}
\end{equation}
For $n=3$ and larger we are not aware of analytic expressions for the symbols $b^{(m_1,\ldots,m_n)}_{l_1,\ldots, l_n}$ but it is easy to determine them numerically from eq.~\eqref{eqc.1} and to tabulate them when needed. 

\section{Impact parameter dependence}
\label{appd}

In this appendix we show that the Bessel-Fourier coefficients of the expectation value of the enthalpy density at fixed reaction plane angle $\phi_R$ as in Eq.\ \eqref{eq3.2} vanish for small impact parameter $b$ like
\begin{equation}
\bar w^{(m)}_l \sim b^{|m|} + {\cal O}(b^{|m|+2}).
\label{eq:ImpactParameterDep1}
\end{equation}
We consider a collision of two (equal size) nuclei with their centers separated by the impact parameter $b$. We choose the coordinate origin to be in the middle of the two nucleus centers. The expectation value for enthalpy can then only depend on the distances from the two centers, $r_A^2 = r^2+b^2/4 + b r \cos(\phi-\phi_R)$ and $r_B^2=r^2+b^2/4-br \cos(\phi-\phi_R)$, or, equivalently on $u=(r_A^2+r_B^2)/2 = r^2+b^2/4$ and $v=(r_A^2-r_B^2) = br \cos(\phi-\phi_R)$. Moreover, symmetry reasoning requires that the expectation value of enthalpy $\bar w$ is a symmetric function of $v$. One can therefore write
\begin{equation}
\bar w(\vec x) = \bar w(u,v) = \sum_{\substack{n=0\\ n\; \text{even}}}^\infty \frac{1}{n!} \bar w^{(0,n)}(u,0) \left[ b\, r \cos(\phi-\phi_R) \right]^{n}.
\label{eq:ImpactParameterDep2}
\end{equation}
In the last step we have expanded in the argument $v$ as one can do at least for small impact parameter $b$. One can now take the Bessel-Fourier transform of this expression. One finds that the coefficients $\bar w^{(m)}_l$ have contributions only from terms on the right hand side of Eq.\ \eqref{eq:ImpactParameterDep2} with $n\geq |m|$. This implies Eq.\ \eqref{eq:ImpactParameterDep1}.

In a similar way, one can write the correlation function in eq.\ \eqref{eq:TwoPointBesselExp} as a function of $u_{1}= r_1^2+b^2/4$,  $u_{2}= r_2^2+b^2/4$, $v_{1}=r_1 b \cos(\phi_1-\phi_R)$ and $v_{2}=r_2 b \cos(\phi_2-\phi_R)$. Symmetry reasons require that this is a symmetric function under $v_1\to-v_1$, $v_2\to-v_2$. One can write
\begin{equation}
\begin{split}
& C(r_1,r_2,\phi_2,\phi_2) = C(u_1,u_2,v_1,v_2) \\
 & = \sum_{\substack{n_1,n_2=0\\ n_1+n_2\; \text{even}}}^\infty \frac{1}{n_1! n_2!} C^{(0,0,n_1,n_2)}(u_1,u_2,0,0) \left[ b r_1 \cos(\phi_1-\phi_R) \right]^{n_1} \left[ b r_2 \cos(\phi_2-\phi_R) \right]^{n_2}. 
\end{split}
\end{equation}
When one expands this into a Fourier series one finds that for small $b$ one has
\begin{equation}
C^{(m_1,m_2)}_{l_1,l_2} \sim b^{|m_1|+|m_2|} + {\cal O}(b^{|m_1|+|m_2|+2}).
\end{equation}
This implies in particular that $C^{(m,m)}_{l_1,l_2} \sim b^{2|m|} + {\cal O}(b^{2|m|+2})$.

\end{appendix}

\acknowledgments
We acknowledge useful discussions with A.~Beraudo, M.~Martinez Guerrero, U.~Heinz, 
J.Y.~Ollitrault, D.~Teaney, N.~Tetradis, L.~Yan and K.~Zapp.

\end{document}